\documentclass[12pt,times]{article}
\pdfoutput=1
 \usepackage[pdftex]{color,graphicx}
\usepackage{tocbibind}
\usepackage{cite}
\usepackage{enumerate}
\usepackage{amsmath}
\usepackage{amssymb}
\usepackage{alltt}
\usepackage{setspace}
\usepackage{subfigure}
\usepackage{sidecap}
\usepackage{sectsty}
\usepackage{wasysym}
\partfont{\large}
\sectionfont{\large}
\subsectionfont{\small}
\usepackage[left=2cm,top=2cm,right=3cm,head=1cm,foot=1cm]{geometry}
\usepackage{mathrsfs}
\usepackage[T1]{fontenc}
\usepackage{authblk}
\numberwithin{equation}{section}
\let\oldsqrt\sqrt
\def\sqrt{\mathpalette\DHLhksqrt}
\def\DHLhksqrt#1#2{%
\setbox0=\hbox{$#1\oldsqrt{#2\,}$}\dimen0=\ht0
\advance\dimen0-0.2\ht0
\setbox2=\hbox{\vrule height\ht0 depth -\dimen0}%
{\box0\lower0.4pt\box2}}
\newcommand{\al}{\alpha}
\newcommand{\g}{\gamma}
\newcommand{\e}{\varepsilon}
\newcommand{\ta}{\theta}
\newcommand{\z}{\zeta}

\newcommand{\ph}{\varphi}

\newcommand{\pa}{\partial}
\newcommand{\na}{\nabla}

\newcommand{\ld}{\lambda}

\newcommand{\ml}{\left(\begin{matrix}}
\newcommand{\mr}{\end{matrix}\right)}

\newcommand{\w}{\omega}
\newcommand{\W}{\Omega}
\newcommand{\Z}{\mathcal Z}

\newcommand{\op}{\mathcal O}
\newcommand{\del}{\delta}
\newcommand{\Del}{\Delta}

\newcommand{\re}{\text{Re}}
\newcommand{\im}{\text{Im}}

\newcommand{\half}{\tfrac{1}{2}}
\newcommand{\third}{\tfrac{1}{3}}
\newcommand{\fourth}{\tfrac{1}{4}}
\newcommand{\sixth}{\tfrac{1}{6}}

\newcommand{\s}{\sigma}

\newcommand{\Ds}{\mathscr D}

\newcommand{\varta}{\vartheta}

\newcommand{\Cs}{\mathscr C}

\newcommand{\THORN}{\text{\th}}
\newcommand{\ETH}{\text{\dh}}

\begin{document}
\title{Gravitational shockwaves on rotating black holes}
\date{}
\author{Yoni BenTov\thanks{Current affiliation: Perimeter Institute for Theoretical Physics, Waterloo, ON N2L 2Y5, Canada} \\ \small Institute for Quantum Information and Matter,\\ \small California Institute of Technology, Pasadena, CA 91125 \and Joe Swearngin\\ \small Department of Physics,\\ \small University of California, Santa Barbara, CA 93106}
\maketitle
\begin{abstract}
We present an exact solution of Einstein's equation that describes the gravitational shockwave of a massless particle on the horizon of a Kerr-Newman black hole. The backreacted metric is of the generalized Kerr-Schild form and is Type II in the Petrov classification. We show that if the background frame is aligned with shear-free null geodesics, and if the background Ricci tensor satisfies a simple condition, then all nonlinearities in the perturbation will drop out of the curvature scalars. We make heavy use of the method of spin coefficients (the Newman-Penrose formalism) in its compacted form (the Geroch-Held-Penrose formalism). 
\end{abstract}

\tableofcontents

\section{Motivation}

Black holes are thermodynamic systems whose microscopic description we still do not understand. After the original work on black hole thermodynamics by Christodoulou \cite{christodoulou_thermo}, Penrose and Floyd \cite{penrose_floyd}, Carter \cite{carter_thermo}, Bekenstein \cite{bekenstein_thermo}, and Bardeen, Carter, and Hawking \cite{bardeen_carter_hawking}, Hawking justified the analogy between the surface gravity\footnote{We use ``$\alpha$'' instead of the more conventional ``$\kappa$'' for surface gravity because ``$\kappa$'' has been commandeered by Newman and Penrose (see Sec.~\ref{sec:matter and gauge}).} $\al$ and a temperature $T$ by predicting that an isolated black hole will radiate as a black body at the expected temperature $T = \frac{\al}{2\pi}$ \cite{hawking, hawking_2}. About 20 years later, Strominger and Vafa vindicated the analogy between the horizon area $A$ and an entropy $S$ by enumerating microstates in string theory to derive the expected result $S = \fourth A$ for extremal black holes in $4+1$ dimensions \cite{microstates}.
\\\\
We will not recount the subsequent history of microstate counting. Suffice it to say that the calculations from string theory, while eminently laudable, are restricted to black holes near extremality and may not provide enough insight into the statistical mechanics behind the conventional black holes of general relativity for generic values of their parameters. It would be helpful to establish a complementary strategy for black hole statistical mechanics tailored to an expansion around the Schwarzschild solution. 
\\\\
One such alternative is the S-matrix approach of 't~Hooft \cite{t_hooft_interpretation, t_hooft_S_matrix}. Motivated by this and by Shenker and Stanford's investigation of the butterfly effect \cite{shenker_stanford, stringy_effects}, Kitaev recently proposed a quantum field theory in $0+1$ dimensions \cite{kitaev} whose low-energy effective action is that of dilaton gravity in $1+1$ dimensions \cite{maldacena_stanford_yang, kitaev_suh}. Details of this model were explored further by Maldacena and Stanford \cite{maldacena_stanford}. Since the equations of motion derived from the effective action admit the AdS$_2$ black hole as a solution \cite{AdS2_gravity}, Kitaev's calculation demonstrates that the thermodynamic limit of a quantum mechanical model\footnote{As remarked by Witten, ``the average of a quantum system over quenched disorder is not really a quantum system'' \cite{witten_syk}. Strictly speaking it is only a quantum mechanical model if the average captures the physics of a single realization with fixed couplings $J_{jk\ell m}$. We thank Yonah Lemonik for a discussion about this important point.} can produce a bona fide black hole horizon, albeit in lower-dimensional scalar-tensor gravity, not in ($3+1$)-dimensional Einstein gravity. 
\\\\
Foundational to all of this is an exact solution of Einstein's equation that describes the gravitational backreaction of a massless particle on the future horizon of a Schwarzschild black hole: the \textit{Dray-'t~Hooft gravitational shockwave} \cite{shockwave}.\footnote{This solution can be viewed as the generalization of the Aichelburg-Sexl shockwave \cite{flat_shockwave} to curved spacetime or as an application of Penrose's ``scissors-and-paste'' method for gluing together known solutions of Einstein's equation to form new solutions \cite{cut_and_paste}.} 
\\\\
That solution was generalized to the Reissner-Nordstr\"om (RN) black hole by Alonso and Zamorano \cite{alonso_zamorano} and by Sfetsos \cite{sfetsos}, who also adapted the shockwave to other static backgrounds. Kiem, Verlinde, and Verlinde \cite{kiem_verlinde_verlinde} used a perturbative variant of the Dray-'t Hooft result to see how gravitational interactions might affect black hole evaporation. And Polchinski \cite{polchinski_chaos} revisited the solution to refine 't~Hooft's ``relation between a given black hole S-matrix element and another with an additional ingoing particle,'' culminating in a reformulated argument for the firewall \cite{AMPS, polchinski_marolf}.
\\\\
In his exposition of the S-matrix framework, 't Hooft did not concern himself with more general black hole backgrounds, opining that ``[c]onceptually, generalization of everything we say to these cases should be straightforward'' \cite{t_hooft_S_matrix}. Perhaps, but in this paper our principal ambition is to galvanize the search for a statistical mechanics underlying astrophysical black holes \cite{ligo}, whose equilibrium field configurations are described by the Kerr geometry. 
So if we intend to adapt 't~Hooft's blueprint and \-Kitaev's recent insights to the microscopics of rotating black holes, then our very first preliminary step must be to generalize 't~Hooft's formula for the transition amplitude.
\\\\
That is what we do here: We generalize the Dray-'t Hooft gravitational shockwave to the Kerr-Newman background, which is the most general asymptotically flat black hole in four spacetime dimensions. Readers familiar with gravitational shockwaves and the method of spin coefficients could skip to our metric ansatz described by Eqs.~(\ref{eq:shifted tetrad}) and~(\ref{eq:shift ansatz}), and then to our main results: the Ricci tensor in Eq.~(\ref{eq:Ricci tensor from the shift}), the Ricci scalar $\Phi_{22}$ in Eq.~(\ref{eq:beautiful Phi22}), and the differential operator in Eq.~(\ref{eq:result}). We acknowledge that this provides only the most tentative intimation toward a microscopic theory of the Kerr-Newman spacetime, but it is a new exact solution of Einstein's equation and therefore deserves to be studied in its own right.
\\\\
Only late in our venture did we learn that Balasin generalized the Ricci tensor for the Dray-'t Hooft solution with the express aim of including rotation in the formalism \cite{balasin}.\footnote{We found Balasin's paper after we had already computed the Ricci tensor but before we managed to express it in the relatively compact and geometrical form described by Eqs.~(\ref{eq:Ricci tensor from the shift}),~(\ref{eq:beautiful Phi22}), and~(\ref{eq:result}).} But he did not complete the calculation, stating only that ``it would be interesting to apply it to a rotating, i.e. Kerr black hole'' and that ``[w]ork in this direction is currently in progress.'' Similar comments were made by Alonso and Zamorano \cite{alonso_zamorano} and by Taub \cite{taub}. We have not found later articles by any of those authors that contain our results.
\\\\
In Sec.~\ref{sec:kerr-newman}, we review everything required to follow the calculation---those unfamiliar with null frames will likely have to supplement this with standard references like Chandrasekhar \cite{chandrasekhar} and Penrose and Rindler \cite{penrose}. In Sec.~\ref{sec:shifted tetrad} we recast the Dray-'t~Hooft geometry as a shift of the null frame, explain how to include rotation, and compute the spin coefficients for generalized Kerr-Schild metrics. In Secs.~\ref{sec:petrov} and~\ref{sec:shifted ricci} we specialize to shear-free geodesic congruences and compute the shifted curvature scalars. Sec.~\ref{sec:sausage} is where the heavy lifting begins: We engage the rotating shockwave and compute some preliminary identities for derivatives of the shift function. This leads to Sec.~\ref{sec:ricci tensor}, where we complete the calculation and announce the differential equation for the shockwave's angular profile. We offer some closing thoughts in Sec.~\ref{sec:end}, and we explain in the appendix how to change metric signature from mostly-minus to mostly-plus.

\section{The Kerr-Newman black hole}\label{sec:kerr-newman}

To enable the reader to work through this document, we will first describe the Kerr-Newman black hole using the method of spin coefficients. 
\\\\
This method was invented by Newman and Penrose (NP) \cite{NP} and refined into a ``compacted'' version by Geroch, Held, and Penrose (GHP) \cite{GHP}, a refinement that has since fallen by the wayside but that we found indispensable. Beside our primary aim of generalizing the gravitational shockwave, our secondary aim is to provide a detailed example of how to use the formalism. As far as rotating black holes are concerned, the flip side of the method of spin coefficients is the madness without it.

\subsection{Null frame}

Our account of the spacetime will begin with a collection of \textit{frame field 1-forms}
\begin{equation}\label{eq:form convention}
e^a \equiv e_\mu^{\,a}\,dx^\mu \equiv (-l',\,-l,\,m',\,m)\;,
\end{equation}
in terms of which the line element is
\begin{equation}\label{eq:line element}
ds^2 = -2ll'+2mm'\;. 
\end{equation}
A tactical advantage of deploying a frame formulation is to never have to look at a line element, so we will not show $ds^2$ explicitly---we will always work directly with the frame. To gain our footing we will start with the ``Schwarzschild-like'' coordinates $(t,r,\ta,\ph)$ of Boyer and Lindquist \cite{boyer_lindquist}, which are applicable outside the black hole.
\\\\
Kerr-Newman black holes have a mass $M$, a charge $Q$, and an angular momentum $J$. It is customary to trade $J$ for the ratio $a \equiv J/M$ and to define the ``horizon function'' \cite{chandrasekhar}
\begin{equation}\label{eq:horizon function}
\Del \equiv r^2-2Mr+a^2+Q^2 \equiv (r-r_+)(r-r_-)\;.
\end{equation}
The inner horizon $r_- \equiv M-\sqrt{M^2-a^2-Q^2}$ and the outer horizon $r_+ \equiv M+\sqrt{M^2-a^2-Q^2}$ are defined as the solutions to $\Del = 0$. It is useful to note that $M = \half(r_+ + r_-)$ and $||(a,Q)|| \equiv (a^2+Q^2)^{1/2} = (r_+ r_-)^{1/2}$. 
\\\\
We will be concerned exclusively with the region $r \geq r_+$, so when we refer to ``the'' horizon, we will always mean the outer one. 
\\\\
Since time immemorial Newman has emphasized that rotating black holes are ``complex translations'' of nonrotating ones \cite{newman_janis}. Regardless of whether that means anything, it is convenient to define the complex functions
\begin{equation}
R \equiv r+ia\cos\ta\;,\;\; R_0 \equiv r+ia\;.
\end{equation}
In the above notation, the following null 1-forms describe the Kerr-Newman black hole:
\begin{align}\label{eq:forms}
&l = -dt+\frac{|R|^2}{\Del} dr+a\,\sin^2\ta\,d\ph\;,\;\;l' = \frac{\Del}{2|R|^2}\left( -dt-\frac{|R|^2}{\Del}dr+a\,\sin^2\ta\,d\ph\right)\;,\nonumber\\
&m = \frac{1}{R\sqrt 2}\left( |R|^2\,d\ta+i|R_0|^2\sin\ta\,d\ph-ia\,\sin\ta\,dt\right)\;,\;\; m' = m^*\;.
\end{align}
Given those 1-forms, we solve the matrix inversion problem
%
%
\begin{equation}\label{eq:matrix inversion}
e_\mu^{\,a} e_a^{\,\nu} \equiv \del_\mu^{\,\nu}\;,\;\; e_a^{\,\mu} e_\mu^{\,b} \equiv \del_a^{\,b}
\end{equation}
for the vectors $e_a^{\,\mu}\equiv(l^\mu, l'^\mu, m^\mu, m'^\mu)$. By royal mandate we then introduce the \textit{Newman-Penrose directional derivatives}:
\begin{equation}\label{eq:covariant NP derivatives}
D \equiv l^\mu \na_\mu\;,\;\; D' \equiv l'^\mu \na_\mu\;,\;\; \del \equiv m^\mu\na_\mu\;,\;\; \del' \equiv m'^\mu \na_\mu\;.
\end{equation}
Without loss of generality we can replace the covariant derivatives by partial derivatives and treat the operators $D, D', \del,\del'$ as ordinary vector fields.\footnote{Once the equations of differential geometry are cast in spin coefficient form, all of the dynamical variables will be invariant under coordinate transformations on the base space, thereby becoming scalar fields.} In Schwarzschild-like coordinates, we have:
\begin{align}\label{eq:vectors}
&D = l^\mu \pa_\mu = \frac{|R_0|^2}{\Del}\pa_t+\pa_r+\frac{a}{\Del}\pa_\ph\;,\;\; D' = l'^\mu \pa_\mu = \frac{\Del}{2|R|^2}\left( \frac{|R_0|^2}{\Del}\pa_t-\pa_r+\frac{a}{\Del}\pa_\ph\right)\;,\nonumber\\
&\del = m^\mu \pa_\mu = \frac{1}{R\sqrt 2}\left( \pa_\ta+\frac{i}{\sin\ta}\,\pa_\ph+ia\,\sin\ta\,\pa_t\right)\;,\;\; \del' = \del^*.
\end{align}
We will refer to the forms in Eq.~(\ref{eq:forms}) and the vectors in Eq.~(\ref{eq:vectors}) as the ``standard'' frame. Its ubiquity derives from its utility: It is a principal basis (see Sec.~\ref{sec:gravitational compass}) whose outgoing and ingoing null congruences are geodesic, twisting, and shear-free [see Eq.~(\ref{eq:spin coefficients for kerr-newman})]. Students acquainted with Reissner-Nordstr\"om but hesitant about Kerr-Newman should fiddle with the standard frame until the geometry feels less foreign. 

\subsection{Spin coefficients}

There are two ways to express the classical field theory of gravity, distinguished by whether local invariance under $SO(3,1)$ is \textit{imposed} or \textit{inferred}. Drastically oversimplifying a complicated history, we will say that the former is Cartan's approach, while the latter is Einstein's.\footnote{Penrose and Rindler \cite{penrose} refer to what we call ``Cartan's approach'' as the ``Einstein-Cartan-Sciama-Kibble theory'' (see their Sec.~4.7).}
\\\\
We favor the former. First introduce a frame $e_\mu^a$ and demand invariance of the action under local $SO(3,1)$ transformations:
\begin{equation}\label{eq:transformation for frame}
e^a(x) \to O^a_{\;\;b}(x)\,e^b(x)\;,\;\; O^a_{\;\;c}(x)\, O^b_{\;\;d}(x)\; \eta_{ab} \equiv \eta_{cd}\;.
\end{equation}
Then introduce an $SO(3,1)$ gauge field $\w^a_{\;\;b}$, called the spin connection, to turn ordinary derivatives into covariant derivatives. As for any nonabelian gauge field, the required transformation law is
\begin{equation}
\w^a_{\;\;b}(x) \to O^a_{\;\;c}(x) \left( \del^c_{\;\;d}\,d+ \w^c_{\;\;d}\right) (O^{-1})^d_{\;\;b}(x)\;.
\end{equation}
By birthright the spin connection is antisymmetric: 
\begin{equation}
\w_{ab} = -\w_{ba}\;. 
\end{equation}
The variables $e^a(x)$ and $\w^a_{\;\;b}(x)$ are the independent classical fields in the action. Because we find it productive to work entirely within the internal space, we follow Newman and Penrose and define the \textit{spin coefficients} \cite{NP}
\begin{equation}\label{eq:NP definition of gamma}
\g_{abc} \equiv (\w_\mu)_{ab}\,e_c^{\,\mu}\;.
\end{equation}
Varying the action with respect to the spin connection in a world without fermions implies the torsion-free condition 
\begin{equation}\label{eq:torsion-free}
de_a = \g_{abc}\,e^b\wedge e^c\;.
\end{equation}
Solving this gives the spin coefficients in terms of the frame:
\begin{equation}\label{eq:definition of gamma}
\g_{abc} = \half(\ld_{abc}\!+\!\ld_{cab}\!-\!\ld_{bca})\;,\;\; \ld_{abc} \equiv -(e_a^\mu e_c^\nu-e_c^\mu e_a^\nu)\pa_\mu e_{b\nu}\;.
\end{equation}
While this expression is standard, the path to it depends on one's taste in formalism. 

\subsection{Partial gauge fixing}\label{sec:GHP group}

After Newman and Penrose invented the method of spin coefficients, Geroch, Held, and Penrose recognized that specifying a frame $e_a^{\,\mu} = (l^\mu, l'^\mu, m^\mu, m'^\mu)$ that satisfies the normalization conditions in Eq.~(\ref{eq:matrix inversion}) only partially fixes the gauge in $SO(3,1)$.
\\\\
The remaining ambiguity comprises a boost along the outgoing congruence, the corresponding inverse boost along the ingoing congruence, and a rotation of the transverse plane:
\begin{equation}\label{eq:GHP}
l^\mu \to r(x)\, l^\mu\;,\;\; l'^\mu \to \frac{1}{r(x)}\,l'^\mu\;,\;\; m^\mu \to e^{\,i\varta(x)}\, m^\mu\;,\;\; m'^\mu \to e^{-i\varta(x)}\,m'^\mu\;.
\end{equation} 
We will say that this transformation generates the \textit{GHP group}. It is convenient to define the complex function 
\begin{equation}\label{eq:GHP parameter polar form}
\ld \equiv r^{1/2}e^{\,i\varta/2}
\end{equation}
and to rewrite Eq.~(\ref{eq:GHP}) as
\begin{equation}\label{eq:GHP group}
l^\mu \to \ld\ld^*\, l^\mu\;,\;\; l'^\mu \to \ld^{-1} \ld^{*\,-1}\,l'^\mu\;,\;\; m^\mu \to \ld \ld^{*\,-1}\, m^\mu\;,\;\; m'^\mu \to \ld^{-1}\ld^*\,m'^\mu\;.
\end{equation}
We will say that a function $f_{h,\bar h}$ transforms as the representation\footnote{The bar is part of the name of the weight and does not denote any sort of conjugation. } $(h,\bar h)$ of the GHP group if its transformation law under Eq.~(\ref{eq:GHP group}) has the form:
\begin{equation}\label{eq:GHP law}
f_{h,\bar h} \to \ld^{2h}\ld^{*\,2\bar h}f_{h,\bar h}\;.
\end{equation}
As shorthand for this, we will use the standard notation of representation theory:
\begin{equation}\label{eq:GHP representation}
f_{h,\bar h} \sim (h,\bar h)\;.
\end{equation}
The numbers $(h,\bar h)$ are called the weights\footnote{Penrose and Rindler define $p \equiv 2h$ and $q \equiv 2\bar h$. Either way, the ``boost weight'' and the ``spin weight'' are defined as $\half(p+q) = h+\bar h$ and $\half(p-q) = h-\bar h$ respectively \cite{GHP}. 
} of the function $f_{h,\bar h}$, and such a function is accordingly said to be ``weighted.'' Borrowing group-theoretic jargon from field theory, we will say that weighted quantities \textit{transform as matter fields}. An object that cannot be assigned a transformation law of the form in Eq.~(\ref{eq:GHP law}) for any values of $(h,\bar h)$ will be called ``nonweighted.''\footnote{Something invariant under Eq.~(\ref{eq:GHP law}) is considered to be weighted with weight zero, not nonweighted.} In the language of Eq.~(\ref{eq:GHP representation}), we summarize Eq.~(\ref{eq:GHP group}) as
\begin{equation}
l^\mu \sim (\half,\half)\;,\;\; l'^\mu \sim (-\half,-\half)\;,\;\; m^\mu \sim (\half,-\half)\;,\;\; m'^\mu \sim (-\half,\half)\;.
\end{equation}
Manifest covariance under the GHP group is what defines the compacted formalism: All explicitly written quantities transform according to Eq.~(\ref{eq:GHP law}) for some values of $h$ and $\bar h$. Only objects with the same weights can be added, and the weights of a product of objects are the sums of the weights of each object: 
\begin{equation}
f_{h_1,\bar h_1} \sim (h_1,\bar h_1)\;,\;\; g_{h_2,\bar h_2} \sim (h_2,\bar h_2) \implies f_{h_1,\bar h_1} g_{h_2,\bar h_2} \sim (h_1+h_2,\bar h_1+\bar h_2)\;.
\end{equation}
From Eq.~(\ref{eq:GHP law}) we deduce that complex conjugation exchanges the weights:
\begin{equation}\label{eq:conjugation exchanges weights}
f_{h,\bar h} \sim (h,\bar h) \implies (f_{h,\bar h})^* \sim (\bar h, h)\;.
\end{equation}
Beside complex conjugation, there are two discrete transformations under which the compacted formalism is covariant. The first is the \textit{priming} transformation, which is defined to exchange primed and unprimed quantities:
\begin{equation}\label{eq:priming}
l_\mu \leftrightarrow l'_\mu\;,\;\; m_\mu \leftrightarrow m'_\mu\;.
\end{equation}
In this way the notation from Eq.~(\ref{eq:form convention}) becomes an operation. From Eq.~(\ref{eq:GHP group}) we deduce that priming flips the signs of the weights:
\begin{equation}\label{eq:priming flips weights}
f_{h,\bar h} \sim (h,\bar h) \implies (f_{h,\bar h})' \sim (-h,-\bar h)\;.
\end{equation}
The second discrete transformation is the \textit{Sachs operation}, which is an analog of Hodge duality:
\begin{equation}\label{eq:sachs}
(l_\mu, l'_\mu, m_\mu, m'_\mu) \to (m_\mu, -m'_\mu, -l_\mu, l'_\mu)\;.
\end{equation}
Unlike priming, the Sachs operation does not commute with complex conjugation. It is extremely convenient to streamline the spin coefficient formalism by using a notation that is manifestly covariant under priming. The Sachs operation will instead help us establish geometrical meaning.

\subsection{Matter fields and gauge fields}\label{sec:matter and gauge}
Based on their behavior under Eq.~(\ref{eq:GHP group}), the 12 independent $\g_{abc}$ fall naturally into three sets: weighted quantities associated with $l_\mu$, weighted quantities associated with $l'_\mu$, and nonweighted quantities that transform as gauge fields.
\\\\
The weighted spin coefficients associated with $l_\mu$, along with their weights, are
\begin{equation}\label{eq:weighted spin coefficients for l}
\kappa \equiv \g_{311} \sim \left( \tfrac{3}{2},\,\tfrac{1}{2}\right),\; \tau \equiv \g_{312} \sim \left(\half,\,-\half\right),\;\s \equiv \g_{313} \sim \left(\tfrac{3}{2},\,-\half\right),\; \rho \equiv \g_{314} \sim \left(\half,\,\half\right)\;.
\end{equation}
The weighted spin coefficients associated with $l'_\mu$ are defined by priming, 
which flips the signs of the weights:\footnote{Priming acts on the $SO(3,1)$ indices by exchanging $1\leftrightarrow2$ and $3\leftrightarrow4$. Complex conjugation leaves $1$ and $2$ fixed while exchanging $3\leftrightarrow 4$.}
\begin{equation}\label{eq:weighted spin coefficients for l'}
\kappa' \equiv \g_{422} \sim \left(-\tfrac{3}{2},\,-\half\right),\; \tau' \equiv \g_{421} \sim \left(-\half,\,\half\right),\; \s' \equiv \g_{424} \sim \left(-\tfrac{3}{2},\,\half\right),\; \rho' \equiv \g_{423} \sim \left(-\half,\,-\half\right)\;.
\end{equation}
The gauge fields of the spin coefficient formalism are defined as
\begin{align}\label{eq:gauge fields}
&\e \equiv \half(-\g_{121}\!+\!\g_{341}),\; \beta \equiv \half(-\g_{123}\!+\!\g_{343}),\;\e' \equiv \half(-\g_{212}\!+\!\g_{432}),\; \beta' \equiv \half(-\g_{214}\!+\!\g_{434})\;.
\end{align}
These are gauge fields in the sense that they combine with the NP derivatives of Eq.~(\ref{eq:vectors}) to form weighted derivatives:
\begin{align}\label{eq:GHP covariant derivatives}
&\THORN \equiv D+2h\,\e+2\bar h\,\e^*\;,\;\; \ETH \equiv \del+2h\,\beta-2\bar h\,\beta'^*\;,\nonumber\\
&\THORN' \equiv D'-2h\,\e'-2\bar h\,\e'^*\;,\;\; \ETH' \equiv \del'-2h\,\beta'+2\bar h\,\beta^*\;.
\end{align} 
We will refer to the operators $\THORN$, $\THORN'$, $\ETH$, and $\ETH'$ as \textit{GHP-covariant derivatives}. Typically the covariant derivative of a matter field transforms as the same representation as the field itself, but not so here. For a weighted function $f_{h,\bar h} \sim (h,\bar h)$, we have:
\begin{equation}
\THORN f_{h,\bar h} \sim (h+\half,\bar h+\half)\;,\;\;\THORN' f_{h,\bar h} \sim (h-\half,\bar h-\half)\;,\;\; \ETH f_{h,\bar h} \sim (h+\half,\bar h-\half)\;,\;\; \ETH' f_{h,\bar h} \sim (h-\half,\bar h + \half)\;.
\end{equation}
Evidently the covariant derivatives themselves carry charge:
\begin{equation}\label{eq:covariant derivatives are charged}
\THORN \sim \left(\half,\half\right)\;,\;\;\THORN'\sim\left(-\half,-\half\right)\;,\;\; \ETH \sim \left(\half,-\half\right)\;,\;\; \ETH' \sim \left(-\half, \half\right)\;.
\end{equation}
%
%
%

\subsection{Null Cartan equations}

Expressed in the NP hieroglyphs of Eqs.~(\ref{eq:weighted spin coefficients for l})-(\ref{eq:gauge fields}), the torsion-free condition of Eq.~(\ref{eq:torsion-free}) becomes four fundamental relations:
\begin{align}\label{eq:null cartan}
&dl = -2\,\re(\e)\,l\wedge l'+2i\,\im(\rho)\,m\wedge m'+\left[\, \left(\tau\!-\!\beta\!+\!\beta'^*\right)\,m'\wedge l+\kappa\, m'\wedge l'+c.c.\,\right]\;,\nonumber\\
&dm = (\beta\!+\!\beta'^*)\,m\wedge m'-(\tau\!-\!\tau'^*)\,l\wedge l'+\left[\, \left( \rho-2i\,\im(\e)\right)\,m\wedge l'+\s\, m'\wedge l'+c.c.'\,\right]\;,
\end{align}
and their primes. 
We will call these the \textit{null Cartan equations}. 
\\\\
By computing the exterior derivatives of the forms in Eq.~(\ref{eq:forms}), arranging them to match the right-hand sides in Eq.~(\ref{eq:null cartan}), and solving the resulting equations, we can find the Kerr-Newman spin coefficients: 
\begin{align}\label{eq:spin coefficients for kerr-newman}
&\kappa = \kappa' = \s = \s' = 0\;,\;\;\rho = \frac{1}{R^*}\;,\;\; \rho' = -\,\frac{\Del}{2|R|^2}\;\frac{1}{R^*}\;,\;\;\tau = \frac{ia\,\sin\ta}{\sqrt 2 |R|^2}\;,\;\; \tau' = \frac{ia\,\sin\ta}{\sqrt2 (R^*)^2}\;,\nonumber\\
&\e = 0\,,\;\e' = \rho'+\frac{2r\!-\!r_+\!-\!r_-}{4|R|^2}\;,\;\;\beta = -\,\frac{\cot\ta}{2\sqrt 2 R}\;,\;\; \beta' = \tau'+\beta^*\;.
\end{align} 
Because of their noncovariance under Eq.~(\ref{eq:GHP group}), the above $\e'$ and $\beta'$ should be understood strictly numerically. Also note that $|\tau|^2 = |\tau'|^2$, which will be useful later.


%
%
%
%
%
%

\subsection{Timelike expansion and timelike twist}\label{sec:timelike expansion and twist}

Every bard recounts legends of refraction ($\kappa$), expansion ($\re\;\rho$), twist ($\im\;\rho$), and shear ($\s$), but nary a soul tells tales of $\tau$.\footnote{Sachs, who pioneered the optical analogy for the spin coefficients, does not explain $\tau$ or $\tau'$ in his original paper \cite{sachs_1961}. Szekeres, in the paper from which we extracted the term ``refraction'' for $\kappa$, calls the spin coefficient $\tau$ (which he denotes $\Omega$) the ``\textit{angular velocity} or \textit{rotation} of the null congruence,'' but he does not explain why \cite{szekeres_refraction}. In a subsequent lecture, Sachs seems to have implicitly recognized this interpretation of $\tau$ by also choosing the symbol $\Omega$ to denote it, but he does not justify the notation \cite{sachs_lecture}. An appraisal of the null Cartan equations within the formal context of lightcone kinematics as originally articulated by Dirac \cite{dirac_lightcone} affirms this interpretation but with $\tau$ and $\tau'$ switched.}
\\\\
%
We would like to elevate the standing of $\tau$ and $\tau'$ to match the renown of their colleagues, because these neglected spin coefficients convey the relativistic effects of rotating bodies at least as directly as $\im(\rho)$ and $\im(\rho')$ do---a cursory assessment of Eq.~(\ref{eq:spin coefficients for kerr-newman}), for instance, reveals the suggestive factor $a\sin\ta$. Our North Star will be the Sachs operation of Eq.~(\ref{eq:sachs}).
\\\\
The combinations $\tau\pm\tau'^*$, rather than $\tau$ and $\tau'$ separately, will appear front and center in the subsequent analysis, so let us consider their meaning and christen them with appropriate names. Sachs conjugation of the expansion and twist provides:
%
\begin{align}
&\re(\rho) \equiv \half(\rho\!+\!\rho^*)\to \half(\tau\!+\!\tau'^*) = -\,\frac{a^2\sin(2\ta)}{2\sqrt 2 |R|^2 R}\;\;, \nonumber\\
&\im(\rho) \equiv \tfrac{1}{2i}(\rho\!-\!\rho^*) \to \tfrac{1}{2i}(\tau\!-\!\tau'^*) = \frac{ra\sin\ta}{\sqrt 2 |R|^2 R}\;\;.\label{eq:tau and tau'}
\end{align}
Consequently, we will refer to $\tau+\tau'^*$ and $\tau-\tau'^*$ as the \textit{timelike expansion} and \textit{timelike twist}.
\\\\
Even though we performed the Sachs operation on spin coefficients associated with $l^\mu$, the result involved both $\tau$ and $\tau'$. While this may be jarring at first sight, GHP covariance requires it: The spin coefficients $\rho$ and $\rho^*$ have the same weights and therefore can be added and subtracted at will, but $\tau$ and $\tau^*$ transform differently under Eq.~(\ref{eq:GHP group}). 
Only $\tau$ and $\tau'^*$ can be added and subtracted. 

\subsection{Kruskal-like coordinates}\label{sec:Kruskal coordinates}

To put all this formalism to work, we will need to forge Kruskal-like coordinates. First recall the known result for the surface gravity:
\begin{equation}\label{eq:surface gravity}
\al = \frac{r_+ - r_-}{2(r_+^2+a^2)}\;.
\end{equation}
With that we define the null coordinates $U$ and $V$ outside the black hole:
\begin{align}
&U \equiv -e^{-\al u}\;,\;\; V \equiv +e^{+\al v}\;,\;\; u \equiv t-r_*\;,\;\; v \equiv t+r_*\;,\;\; dr_* \equiv \frac{|R_0|^2}{\Del}dr\;.
\end{align} 
Note that $U < 0$, which is the standard convention. We choose the integration constant in the tortoise coordinate $r_*$ such that the product of $U$ and $V$ is\footnote{Since we always work with $r > r_-$, we have dropped the absolute values that emerge from integrating $dr_*$. Our coordinates are singular at the inner horizon, and a different set of Kruskal-like coordinates must be established to cross it.}
\begin{equation}\label{eq:UV}
UV = -\;\frac{\Del}{r_+r_-}\left( \frac{r}{r_-}-1\right)^{-k}\,e^{\,2\al r}\;,\;\;k \equiv \frac{r_-^2+a^2}{r_+^2+a^2}+1\;.
\end{equation}
Considered an implicitly defined function of $U$ and $V$, the coordinate $r$ retains its desirable property from the nonrotating case of depending only on the product $UV$. As written in Eq.~(\ref{eq:UV}), the ratio $\frac{\Del}{UV}$ is manifestly finite and nonzero at $r = r_+$:
\begin{equation}\label{eq:Delta/UV}
c \equiv -\left.\frac{\Del}{UV}\right|_{r\,=\,r_+} = r_+ r_-\left( \frac{r_+}{r_-}-1\right)^{k} e^{-2\al r_+}\;.
\end{equation}
For later convenience, we also differentiate both sides of Eq.~(\ref{eq:UV}) and rearrange to solve for the partial derivatives of $r(U,V)$:
\begin{equation}\label{eq:derivatives of r}
U\pa_U r = V\pa_V r = \frac{\Del}{\Del'(r)+\left( 2\al-\frac{k}{r-r_-}\right)\Del}\;.
\end{equation}
For any function $F(r)$ that depends only on the radial coordinate, we therefore have:
\begin{equation}\label{eq:derivative of r at horizon is zero}
U\pa_U F(r) = V\pa_V F(r) = F'(r)\, U\pa_U r\;\;\text{ and }\;\; \left. U\pa_U r\right|_{r\,=\,r_+} = 0\;.
\end{equation}
We will sometimes use a subscript ``+'' to label quantities evaluated at the horizon. For instance, $|R_+|^2 \equiv r_+^2+a^2\cos^2\ta$ and $|R_{0+}|^2 \equiv r_+^2+a^2$. 
\\\\
Finally, we define the delayed angular coordinate and the angular velocity at the horizon:
\begin{equation}\label{eq:chi}
\chi \equiv \ph - \W_H t\;,\;\; \W_H = \frac{a}{r_+^2+a^2}\;.
\end{equation}
%

\subsection{A smooth frame}\label{sec:smooth tetrad}

Smooth coordinates are not enough---we also need a smooth frame. From the standard basis written in Kruskal-like coordinates, we perform the following GHP transformation:
\begin{equation}\label{eq:rescaling}
l_\mu \to \hat l_\mu = -U l_\mu\;,\;\; l'_\mu \to \hat l'_\mu = -U^{-1} l'_\mu\;,\;\; m_\mu \to \hat m_\mu = m_\mu\;.
\end{equation}
This describes the special case 
\begin{equation}
\ld = \ld^* = (-U)^{1/2}
\end{equation}
of the transformation in Eq.~(\ref{eq:GHP group}). A hatted function with weights $(h,\bar h)$ is then related to its unhatted counterpart by
\begin{equation}\label{eq:hatting}
\hat f_{h,\bar h} = (-U)^{h+\bar h}\, f_{h,\bar h}\;.
\end{equation}
The spin coefficients $\rho$ and $\rho'$ in the hatted basis,
\begin{equation}\label{eq:hatted expansions}
\hat \rho = \frac{1}{R^*}(-U)\;\text{ and }\;\; \hat\rho' = \frac{1}{2|R|^2}\left(\frac{\Del}{UV}\right)\frac{1}{R^*}V\;,
\end{equation}
go to zero at the future horizon ($U = 0$) and the past horizon ($V = 0$) respectively. These furnish local definitions for each part of the horizon.
\\\\
Because $\tau \sim (\half,-\half)$ and $\tau' \sim (-\half,\half)$, those two spin coefficients are invariant under the rescaling in Eq.~(\ref{eq:rescaling}):
\begin{equation}
\hat\tau = \tau\;,\;\; \hat\tau' = \tau'\;.
\end{equation}
After changing coordinates from $(t,r,\ta,\ph)$ to $(U,V,\ta,\chi)$ and applying Eq.~(\ref{eq:rescaling}), we obtain the following frame field 1-forms:
\begin{align}\label{eq:kruskal forms}
&\hat l = \frac{-1}{2\al}\left( 1+\frac{|R|^2}{|R_0|^2}-\W_H\, a\sin^2\ta\right)dU-\frac{U}{V}\left( 1-\frac{|R_{0+}|^2}{|R_0|^2}\right)\frac{a^2\sin^2\ta}{2\al |R_{0+}|^2}\; dV-U a\sin^2\ta \;d\chi\;,\nonumber\\
&\hat l' \!=\! \frac{\Del}{2|R|^2}\!\left[ \frac{1}{2\al}\!\left( 1\!+\!\frac{|R|^2}{|R_0|^2}\!-\!\W_H\,a\sin^2\ta\right)\!\frac{\,dV}{UV}+\frac{1}{U^2}\left( 1\!-\!\frac{|R_{0+}|^2}{|R_0|^2}\right)\!\frac{a^2\sin^2\ta}{2\al |R_{0+}|^2}\,dU\!-\!\frac{a\sin^2\ta}{U}\,d\chi\right],\nonumber\\
&\hat m = \frac{1}{R\sqrt 2}\left[ |R|^2\,d\ta+i|R_0|^2\sin\ta\,d\chi+\frac{ia\,\sin\ta}{2\al |R_{0+}|^2}\;\frac{r+r_+}{r-r_-}\;\frac{\Del}{UV}\left( U\,dV-V\,dU\right)\right]\;.
\end{align}
The corresponding directional derivatives are
\begin{align}\label{eq:kruskal vectors}
&\hat D = -2\al\,|R_0|^2\frac{UV}{\Del}\,\pa_V-a\;\frac{U}{\Del}\left( 1-\frac{|R_0|^2}{a}\W_H\right)\,\pa_\chi\;,\nonumber\\
&\hat D' = \frac{\Del}{2|R|^2}\left[ 2\al\,\frac{|R_0|^2}{\Del}\,\pa_U-\frac{a}{U\Del}\left( 1-\frac{|R_0|^2}{a}\W_H\right)\,\pa_\chi\right]\;,\nonumber\\
&\hat\del = \frac{1}{R\sqrt2}\left[ \pa_\ta+\frac{i}{\sin\ta}\,\frac{|R_+|^2}{|R_{0+}|^2}\,\pa_\chi+i\al\,a\sin\ta\left( -U\,\pa_U+V\,\pa_V\right)\right]\;.
\end{align}
We will refer to the forms in Eq.~(\ref{eq:kruskal forms}) and the vectors in Eq.~(\ref{eq:kruskal vectors}) as the ``horizon'' frame (or simply as the ``hatted'' one). Each component of the 1-forms in Eq.~(\ref{eq:kruskal forms}) and of the vectors in Eq.~(\ref{eq:kruskal vectors}) is finite at $U = 0$ for fixed $V$, and at $V = 0$ for fixed $U$.

\subsection{Spacelike and timelike curvatures}\label{sec:submanifolds}

Commutators of covariant derivatives beget curvature. By composing GHP derivatives on a test function $\xi_h \sim (h,0)$, we define the \textit{spacelike and timelike curvatures} $\mathcal K$ and $\mathcal K_s$:
\begin{align}\label{eq:formulas for K and Ks}
&\mathcal K\,\xi_h \equiv  -\frac{1}{2h}\left([\ETH,\ETH']+2i\,\im(\rho)\THORN'-2i\,\im(\rho')\THORN\right)\xi_h\;,\nonumber\\
&\mathcal K_s\,\xi_h \equiv  \frac{1}{2h}\left([\THORN,\THORN']+(\tau\!-\!\tau'^*)\ETH'+(\tau^*\!-\!\tau')\ETH\right)\xi_h\;.
\end{align}
Twice the real part of $\mathcal K$ is the ordinary notion of intrinsic (or ``Gaussian'') curvature in Riemannian geometry. The imaginary part is an extrinsic quantity that we will call the extrinsic curvature.\footnote{This is not to be conflated with what numerical relativists call the extrinsic curvature, which is part of the spin connection. See, for example, the discussion of contorted surfaces on p.~400 of \textit{Spinors and Spacetime} \cite{penrose}.} 
For Kerr-Newman, the intrinsic and extrinsic curvatures are
\begin{align}
\re(\mathcal K) = \frac{1}{2|R|^6}&\left\{ r^2(r^2+a^2)+\left[ (r_+^2+a^2)-4(r_+r+a^2)-(r^2-r_+^2) \right]a^2\cos^2\ta \right. \nonumber\\
&\left. +\,4\,\al\, (r_+^2+a^2) (r-r_+)\, a^2\cos^2\ta   \right\}
\end{align}
and
\begin{align}
\im(\mathcal K) =\frac{a\cos\ta}{|R|^6}&\left\{ (r-r_+)\,r_+\, r + (2a^2+r^2)\,r - r_+ a^2\cos^2\ta  \right. \nonumber\\
&\left. +\, \al (r_+^2+a^2)\left[ (2r_+ - r)\,r+a^2\cos^2\ta\right]\right\}\;. 
\end{align}
%
At the horizon, the intrinsic curvature is \cite{smarr_E3}
\begin{equation}\label{eq:intrinsic curvature at horizon}
\left.\re(\mathcal K)\right|_{r\,=\,r_+} = \frac{|R_{0+}|^2}{2|R_+|^6}(r_+^2-3a^2 \cos^2\ta)\;.
\end{equation}
Only at $r = r_+$ should the denominations ``intrinsic'' and ``extrinsic'' be taken literally, because only there do $m^\mu$ and $m'^\mu$ form a surface. In contrast, the real and imaginary parts of $\mathcal K_s$ can never be interpreted that way, because $l^\mu$ and $l'^\mu$ never form a surface.\footnote{Take the hatted basis and consider the commutators of covariant derivatives on a test function of weight~$(0,0)$: We have $[\hat\ETH,\hat\ETH']=(\hat\rho\!-\!\hat\rho^*)\hat\THORN'-(\hat\rho'\!-\!\hat\rho'^*)\hat\THORN$ and $[\hat\THORN,\hat\THORN'] = (\hat\tau\!-\!\hat\tau'^*)\hat\ETH'+(\hat\tau^*\!-\!\hat\tau')\hat\ETH$. The right-hand side of the former vanishes at $U = V = 0$, while the right-hand side of the latter never vanishes except at the poles. We thank Leo Stein for emphasizing this to us.} 
So we leave the Kerr-Newman timelike curvature as a complex quantity:
\begin{align}
\mathcal K_s &= \frac{-2(3r_+\!-\!2r)rr_+\! +\!2iar_+(4r\!-\!r_+)\!\cos\ta\!+\!a^2[5r\!-\!2r_+\!-\!(2r_+\! -\!r)\!\cos(2\ta)]\!+\!2ia^3\cos^3\ta}{4R^*|R|^4}\nonumber\\
&+\al\, (r_+^2+a^2)\,\frac{(3r_+\! -\! r)r-ia(2r\!-\!r_+)\cos\ta+a^2\cos^2\ta}{R^*|R|^4}\;.
\end{align}
%
Next we will summarize those remaining aspects of curvature that are pertinent but more or less standard.  
 
\subsection{Curvature scalars}

%
%
The Riemann tensor in the NP frame is 
\begin{equation}
R_{abcd} = \pa_c \g_{abd}-\pa_d\g_{abc}-\g_{ab}^{\;\;\;\;e}(\g_{ced}-\g_{dec})+\g_{aec}\g^e_{\;\;bd}-\g_{aed}\g^e_{\;\;bc}\;.
\end{equation}
From the corresponding Ricci tensor, $R_{ab} \equiv R^c_{\;\;acb}$, Newman and Penrose define a traceless matrix
\begin{equation}\label{eq:traceless ricci}
\phi_{ab} \equiv \half (R_{ab}-\fourth \eta_{ab}\, \eta^{cd}R_{cd})\;.
\end{equation}
For spinorial reasons of no concern to us, they then define the \textit{Ricci scalars} as 
\begin{align}\label{eq:Ricci scalars}
&\Phi_{00} \equiv \phi_{11} \sim (1,1)\;,\;\; \Phi_{01} \equiv \phi_{13} \sim (1,0)\;,\;\; \Phi_{02} \equiv \phi_{33} \sim (1,-1)\;,\nonumber\\
&
\Phi_{22} \equiv \Phi_{00}' = \phi_{22}\sim (-1,-1)\;,\;\; \Phi_{21} \equiv \Phi_{01}' = \phi_{24}\sim(-1,0)\;,\;\;\nonumber\\
&\Phi_{20} \equiv \Phi_{02}' = \phi_{44} = \Phi_{02}^* \sim (-1,1)\;,\;\;\Phi_{10} \equiv \Phi_{01}^* = \phi_{14} \sim (0,1)\;,\nonumber\\
&\Phi_{12} \equiv \Phi_{21}^* = \phi_{23}  \sim (0,-1)\;,\;\;\Phi_{11} \equiv \half (\phi_{12}+\phi_{34}) \sim (0,0)\;.
\end{align}
In the notation of the compacted formalism, we have\footnote{The expression for $\Phi_{00}$ is in fact real but not manifestly so.}
\begin{align}
&\Phi_{00} = -\THORN\rho\!-\!\rho^2\!-\!|\s|^2\!+\!\ETH'\kappa+\!\tau'\kappa\!+\!\tau\,\kappa^*\;,\;\;\Phi_{02} = -\ETH\tau\!-\!\tau^2\!-\!\kappa\kappa'^*\!+\!\THORN'\s+\!\rho'\s\!+\!\rho\,\s'^*\;,\label{eq:Phi00}\\
&\Phi_{01} = \half\left[-\THORN\tau+\THORN'\kappa-\ETH\rho+\ETH'\s-(\tau\!-\!\tau'^*)\rho-(\tau^*\!-\!\tau')\s-(\rho\!-\!\rho^*)\tau+(\rho'\!-\!\rho'^*)\kappa\right]\;.\label{eq:Phi01}
\end{align}
The remaining Ricci scalars of nonzero weight can be defined by priming and conjugating the definitions already listed: $\Phi_{22} =\Phi_{00}'$, $\Phi_{21} = \Phi_{01}'$, $\Phi_{10} = \Phi_{01}^*$, $\Phi_{12} = \Phi_{21}^*$, and $\Phi_{20} = \Phi_{02}^*$. Meanwhile, the Ricci scalar of weight $(0,0)$ is defined in terms of the spacelike and timelike curvatures:
\begin{equation}\label{eq:Phi11}
\Phi_{11} = \half\left( \mathcal K-\mathcal K_s-\kappa\kappa'+\tau\tau'-\s\s'+\rho\rho'\right)\;.
\end{equation}
Tradition compels a fanciful notation for a factor times the trace of the Ricci tensor:
\begin{equation}
\Pi \equiv \tfrac{1}{12}(R_{12}-R_{34}) = -\,\tfrac{1}{24} \eta^{ab} R_{ab}\;.
\end{equation}
Because of its role as the gravitational Lagrangian, we refer to this as the \textit{Einstein-Hilbert curvature}. In GHP notation, it reads
\begin{equation}\label{eq:Pi}
\Pi = \sixth\left[ 2\left( \rho\rho'^*-|\tau|^2+\THORN'\rho-\ETH'\tau\right)+\mathcal K+\mathcal K_s-\kappa\kappa'-\tau\tau'+\s\s'+\rho\rho'\right]\;.
\end{equation}
Finally we are left with the completely traceless part of the curvature:  
\begin{equation}\label{eq:C_abcd}
C_{abcd} \equiv R_{abcd}+\eta_{ad}\,\phi_{bc}+\eta_{bc}\,\phi_{ad}-\eta_{ac}\,\phi_{bd}-\eta_{bd}\,\phi_{ac}+2\,(\eta_{ac}\,\eta_{bd}-\eta_{ad}\,\eta_{bc})\,\Pi\;.
\end{equation}
This is the Weyl tensor in the NP frame, and from it Newman and Penrose define the \textit{Weyl scalars}:
\begin{align}\label{eq:Weyl scalars}
&\Psi_0 \equiv C_{1313} \sim (2,0)\;,\;\; \Psi_1 \equiv C_{1312} \sim (1,0)\;,\;\; \Psi_2 \equiv C_{1342} \sim (0,0)\;,\nonumber\\
&\Psi_3 \equiv \Psi_1' = C_{2421} \sim (-1,0)\;,\;\;\Psi_4 \equiv \Psi_0' = C_{2424} \sim (-2,0)\;.
\end{align}
In GHP notation, the first three of these are\footnote{Having defined $\Psi_1$ and $\Phi_{01}$, we can compose $\THORN$ and $\ETH$ on an arbitrarily-weighted test function and deduce the mixed commutator relation
\begin{equation}\label{eq:[thorn,eth]}
[\THORN,\ETH]\!+\!\rho^*\ETH\!+\!\s\ETH'\!-\!\tau'^*\THORN\!-\!\kappa\THORN'=-2h(\rho'\kappa\!-\!\tau'\s\!+\!\Psi_1)-2\bar h(\s'^*\kappa^*\!\!-\!\rho^*\tau'^*\!\!+\!\Phi_{01})\;.
\end{equation}
\vspace{-10pt}
}
\begin{align}
&\Psi_0 =-\left[ \THORN+(\rho\!+\!\rho^*)\right]\s+\left[ \ETH+(\tau\!+\!\tau'^*)\right]\kappa\;,\label{eq:Psi0} \\
&\Psi_1 = \half\left\{ -\THORN\tau+\THORN'\kappa+\ETH\rho-\ETH'\s-(\tau\!-\!\tau'^*)\rho-(\tau^*\!-\!\tau')\s+(\rho\!-\!\rho^*)\tau-(\rho'\!-\!\rho'^*)\kappa\right\}\;, \label{eq:Psi1} \\
&\Psi_2 = \third\left[ \rho\rho'^*-|\tau|^2+\THORN'\rho-\ETH'\tau-\left( \mathcal K+\mathcal K_s\right)-2\kappa\kappa'+\tau\tau'+2\s\s'-\rho\rho'\right]\;.\label{eq:Psi2}
\end{align}
The remaining two are defined by priming.

\subsection{Gravitational compass and Petrov classification}\label{sec:gravitational compass}

Szekeres conjured an elegant theoretical apparatus called the \textit{gravitational compass} to interpret the Weyl scalars \cite{szekeres}. Following his insight, we will say that $\Psi_2$ describes a Coulomb field, $\Psi_4$ describes a transverse outgoing wave, and $\Psi_3$ describes a longitudinal outgoing wave. The primed quantities, $\Psi_0 \equiv \Psi_4'$ and $\Psi_1 \equiv \Psi_3'$, describe the corresponding ingoing waves.\footnote{The Coulomb component is self-prime. We might also suggest an alternative notation to make Szekeres's interpretation manifest: $\Psi_\perp \equiv \Psi_4$, $\Psi_\parallel \equiv \Psi_3$, $\Psi_C \equiv \Psi_2$, $\Psi_\perp' \equiv \Psi_0$, and $\Psi_\parallel' \equiv \Psi_1$.} 
\\\\
The Weyl scalars are not gauge invariant: A local $SO(3,1)$ transformation $e^a \to O^a_{\;\;b}\, e^b$ results in $\Psi_\al \to \sum_{\beta\, =\, 0}^4 Q_{\al\beta}\Psi_\beta$ for some matrix $Q_{\al\beta}$. We can ask how many $\Psi_\al$ can be simultaneously gauged away, and we can classify spacetimes based on the answer. This is Chandrasekhar's \cite{chandrasekhar} account of the Petrov classification \cite{petrov} of the Weyl tensor. A desire to elucidate the physics behind each Petrov type is what drove Szekeres to engineer the gravitational compass. 
\\\\
We will only study two Petrov types: Type D, in which all of the Weyl scalars beside $\Psi_2$ can be gauged away, and Type II, in which all of the Weyl scalars beside $\Psi_2$ and $\Psi_4$ can be gauged away.\footnote{For a Type II spacetime, we can rotate the frame to trade a nonzero $\Psi_4$ for a nonzero $\Psi_3$. This resolves the superficial discrepancy between Chandrasekhar's \cite{chandrasekhar} and Penrose and Rindler's descriptions \cite{penrose}. Szekeres \cite{szekeres} and Griffiths \cite{griffiths} use the terminology of Penrose and Rindler.} Extending the standard terminology slightly beyond its ordinary usage, we will define a \textit{principal frame} as any basis in which as many Weyl scalars as possible for a given geometry are gauged away.
\\\\
The Kerr-Newman black hole is Type D, and its nonzero Weyl scalar is
\begin{equation}\label{eq:Psi2 KN}
\Psi_2=-\,\frac{1}{(R^*)^3}\left( M-\frac{Q^2}{R}\right)\;.
\end{equation}
Because it carries charge, this black hole is not a vacuum solution---the Weyl scalars are no longer the whole story. Local sources of energy induce Ricci curvature, and in this case the electromagnetic field induces 
\begin{equation}\label{eq:Phi11 KN}
\Phi_{11} = \frac{Q^2}{2|R|^4}\;.
\end{equation}
%

\subsection{Energy scalars}\label{sec:energy scalars}

In the relativistic zeitgeist, the Ricci scalars are considered a stand-in for the energy tensor by means of Einstein's equation. But we find this confusing and will briefly suggest a refined presentation.
\\\\
To match Penrose's traceless Ricci tensor from Eq.~(\ref{eq:traceless ricci}), we define a traceless energy tensor
\begin{equation}
\mathcal T_{ab} \equiv \half \left( T_{ab} - \fourth \eta_{ab}\, \eta^{cd} T_{cd} \right)\;.
\end{equation}
From that, we define ``energy scalars'' analogously to the Ricci scalars: $t_{00} \equiv 8\pi \mathcal T_{11}$, and so on, such that Einstein's equation becomes
\begin{equation}\label{eq:einstein eq}
\Phi_{ij} = t_{ij}\;\;\text{ and }\;\; \Pi = t_\Pi\;\;,
\end{equation} 
with $i,j\in\{0,1,2\}$. For the Kerr-Newman solution, the only nonzero entry is $t_{11}$, which can be expressed in terms of a complex number $\ph_1$ called a Maxwell scalar:\footnote{The Maxwell scalars are defined as the components of the electromagnetic curvature contracted with the vectors of the null frame: 
\begin{equation*}
\ph_0 \equiv F_{\mu\nu}\, l^\mu m^\nu\;,\;\;\ph_1 \equiv \half F_{\mu\nu}\,(l^\mu l'^\nu+m'^\mu m^\nu)\;,\;\; \ph_2 \equiv F_{\mu\nu}\,m'^\mu l'^\nu\;.
\end{equation*}
Note that because $F_{\mu\nu} = -F_{\nu\mu}$, we have $\ph_2 = -\ph_0'$ and $\ph_1' = -\ph_1$. 
}
\begin{equation}\label{eq:phi1 KN}
t_{11} = |\ph_1|^2\;,\;\; \ph_1 = \frac{Q}{\sqrt 2(R^*)^2}\;.
\end{equation}
Our point is that the equation $t_{11} = |\ph_1|^2$ is the statement $T_{\mu\nu} = F_{\mu\rho}F_\nu^{\;\rho}-\fourth g_{\mu\nu} F_{\rho\s}F^{\rho\s}$ in the internal space, and the equation $\Phi_{11} = t_{11}$ is Einstein's equation in the internal space. The typically stated relation $\Phi_{11} = |\ph_1|^2$ combines both.
\\\\
Having traipsed through the background geometry, we are now ready to perturb it.

\section{Shifted frame and Kerr-Schild form}\label{sec:shifted tetrad}

Relative to the standard frame of the Kerr-Newman background, and in terms of a general function $S(t,r,\ta,\ph)$, we define the \textit{shifted frame}\footnote{Only during revisions did we find the apropos work by Fels and Held \cite{rides_again}. While their shift is like ours, their analysis differs. Strikingly, they consider shifting Type D backgrounds but conclude that ``as seeds they are not very fruitful.'' We disagree.}
\begin{equation}\label{eq:shifted tetrad}
\tilde l \equiv l\;,\;\; \tilde l' \equiv l'+S\,l\;,\;\; \tilde m \equiv m\;.
\end{equation}
It cannot be emphasized enough that the meaning of $\tilde l'$ in components is 
\begin{equation}
\tilde l' = \tilde l'_\mu\, dx^\mu = (l'_\mu + S l_\mu)\, dx^\mu\;,
\end{equation}
not $\tilde l'_\mu\, d\tilde x^\mu$ for some shifted coordinate basis $d\tilde x^\mu$. Otherwise the shift would describe a change of coordinates, not a physical perturbation.
\\\\
Recalling Eq.~(\ref{eq:line element}), we define the shifted line element as
\begin{equation}
d\tilde s^2 \equiv -2\tilde l\tilde l'+2\tilde m\tilde m' = ds^2-2Sll\;.
\end{equation}
Since we have chosen $l^\mu$ to be tangent to a shear-free geodesic congruence of the unshifted spacetime, the shifted line element is of the generalized Kerr-Schild form, as defined by Taub \cite{taub}. If we turn off the angular momentum and the charge and choose the ansatz
\begin{equation}\label{eq:nonrotating ansatz}
S = \frac{\Del}{2r^2}\,\frac{U}{V}\,\del(U)\,f(\ta,\ph)\qquad(a = Q = 0)
\end{equation}
then we will reproduce exactly the Dray-'t Hooft metric \cite{shockwave}. If we turn off the angular momentum but leave the charge nonzero and use the same functional form for the ansatz, we will reproduce the metric of Alonso-Zamorano \cite{alonso_zamorano} and Sfetsos \cite{sfetsos}.

\subsection{From Reissner-Nordstr\"om to Kerr-Newman}

To generalize to a rotating background, we will scrutinize the factors that appear in Eq.~(\ref{eq:nonrotating ansatz}). 
\\\\
First, by revisiting our conventions for the unshifted frame and staring at the definition of the shifted one, we conclude that the factor $\frac{\Del}{2r^2}$ compensates for the asymmetric normalization of $l_\mu$ relative to $l'_\mu$. So the generalization of this factor to the rotating case is clear: 
\begin{equation}
\frac{\Del}{2r^2} \to \frac{\Del}{2|R|^2}\;.
\end{equation}
Second, we have defined the Kruskal-like coordinates so that they mimic the coordinates in the nonrotating case: The future horizon is still at $U = 0$, and the radial function $r$ depends only on the product $UV$. So we might hope that the factor $\frac{U}{V}\,\del(U)$ could remain unmodified.
\\\\
Third, we recognize that the function $f(\ta,\ph)$ is defined only at the origin of Kruskal-like coordinates $(U = V = 0)$. Extrapolating to the Kerr-Newman spacetime should therefore entail the generalization 
\begin{equation}
(\ta,\ph) \to (\ta,\chi)\;.
\end{equation}
This cross-examination of the Dray-'t~Hooft solution coupled with the clear geometrical underpinning of the Newman-Penrose formalism led us to the conviction that the perturbed Kerr-Newman geometry should be described by the shifted frame in Eq.~(\ref{eq:shifted tetrad}) with the following ansatz:
\begin{equation}\label{eq:shift ansatz}
S = \frac{\Del}{2|R|^2}\,\frac{U}{V}\,\del(U)\,f(\ta,\chi)\;.
\end{equation}
We will call $S$ the \textit{shift function}, and we will call $f(\ta,\chi)$ the \textit{horizon field}. When we calculate the curvature scalars, we will work directly with the rescaled frame in Eq.~(\ref{eq:kruskal vectors}), thereby enlisting the rescaled shift function
\begin{equation}\label{eq:hatted shift ansatz}
\hat S = (-U)^{-2}S = \frac{1}{2|R|^2}\frac{\Del}{UV}\,\del(U) f(\ta,\chi)\;.
\end{equation}
Like everything else in the hatted basis, this shift function is finite at the horizon.
\\\\
By comparing the GHP representations $l'_\mu \sim (-\half,-\half)$ and $l_\mu \sim (+\half,+\half)$ in the context of Eq.~(\ref{eq:shifted tetrad}), we deduce that the shift function must transform as
\begin{equation}
S \sim (-1,-1)\;.
\end{equation}
When interpreting the formulas Eqs.~(\ref{eq:shift ansatz}) and~(\ref{eq:hatted shift ansatz}) in the GHP formalism, we assign the horizon field $f(\ta,\chi)$ the weights of the shift function:
\begin{equation}\label{eq:weights of horizon field}
f(\ta,\chi) \sim (-1,-1)\;.
\end{equation}
The remaining factors are to be treated as ordinary functions, not physical degrees of freedom, and are therefore assigned weights $(0,0)$.
\\\\
By explicit calculation, we will indeed find that the ansatz in Eq.~(\ref{eq:shift ansatz}) results in a shifted Ricci tensor of the form 
\begin{equation}
\tilde R_{\mu\nu} = R_{\mu\nu}+R^{\text{ shift}}_{UU}\, \del_\mu^{\,U}\del_\nu^{\,U}
\end{equation}
and therefore correctly generalizes the Dray-'t Hooft solution to a rotating background.

\subsection{Preliminary commentary}

Before focusing on $R_{UU}$, we wish to preview a miracle: If the unshifted frame is aligned with shear-free null geodesics $(\kappa = \s = \kappa' = \s' = 0)$ and if the unshifted $\Phi_{00}$ is zero, the shifted Ricci tensor will depend only \textit{linearly} on the shift function $S$.\footnote{This was in fact noticed by Taub \cite{taub} and by Alonso and Zamorano \cite{alonso_zamorano}.}
\\\\
We will proceed step by step through the spin coefficient formalism to understand why this happens. A practical reason is to derive master formulas for the spin coefficients and curvature scalars of generalized Kerr-Schild spacetimes. For the spin coefficients we will maintain full generality in the background, but for the curvature scalars we will restrict to shear-free geodesic congruences.

\subsection{Shifted spin coefficients}

By shifting both sides of the null Cartan equations [Eq.~(\ref{eq:null cartan})] and solving them, we can express the shifted spin coefficients in terms of their unshifted values. 
\\\\
Start with the equation for $dl$, and tilde every term:\footnote{Note that, in keeping with our advisory remark below the definition of the shifted frame [Eq.~(\ref{eq:shifted tetrad})], we do not tilde the exterior derivative operator.}
\begin{equation}\label{eq:dl tilde}
d\tilde l = -2\,\re(\tilde\e)\,\tilde l\wedge \tilde l'+2i\,\im(\tilde\rho)\,\tilde m\wedge \tilde m'+\left[\, \left(\tilde\tau-\tilde\beta+\tilde\beta'^*\right)\,\tilde m'\wedge \tilde l+\tilde \kappa\, \tilde m'\wedge \tilde l'+c.c.\,\right]\;.
\end{equation}
%
By inserting into the right-hand side the definition of the shifted frame in terms of the unshifted frame and recalling that $l\wedge l = 0$, we find
\begin{equation}
d\tilde l = -2\,\re(\tilde\e)\, l\wedge l'+2i\,\im(\tilde\rho)\, m\wedge m'+\left[ \left(\tilde\tau-\tilde\beta+\tilde\beta'^*+\tilde\kappa\, S\right)\, m'\wedge l+\tilde\kappa\, m'\wedge l'+c.c.\right]\;.
\end{equation}
Since $\tilde l= l$, we have $d\tilde l = dl$, so the left-hand side can be replaced with the untilded version of Eq.~(\ref{eq:dl tilde}). The four basis 2-forms $l\wedge l'$, $m\wedge m'$, $m'\wedge l$, and $m'\wedge l'$ are linearly independent, so we can match their coefficients on both sides to obtain the first set of shifted spin coefficient equations:
\begin{equation}\label{eq:first set of equations}
\re(\tilde\e)= \re(\e)\;,\;\; \im(\tilde\rho) = \im(\rho)\;,\;\; \tilde\tau-\tilde\beta+\tilde\beta'^*+\tilde\kappa\, S = \tau-\beta+\beta'^*\;,\;\; \tilde \kappa = \kappa\;.
\end{equation}
Next up, $dl'$. The right-hand side parallels that for $dl$, 
%
%
%
%
but since $l' = l'+Sl$ the left-hand side is more complicated. Not only do we require the untilded equations for both $dl'$ and $dl$, we also require the exterior derivative of the shift function:
\begin{equation}\label{eq:dS}
dS \equiv dx^\mu\, \pa_\mu S = e^a\,\pa_a S = -l\, D'S-l' DS+m\, \del'S+m'\del S\;.
\end{equation}
%
%
%
Matching the coefficients of the basis 2-forms gives the second set of shifted spin coefficient equations:
\begin{align}\label{eq:second set of equations}
&\re(\tilde \e') = \re(\e')-\re(\e)S+\half DS\;,\;\; \im(\tilde\rho')=\im(\rho')-\im(\rho) S\;,\nonumber\\
&\tilde\tau'\!-\!\tilde\beta'\!+\!\tilde\beta^* = \tau'\!-\!\beta'\!+\!\beta^*\!+\!\kappa^*S\;,\;\;\tilde\kappa'+(\tilde\tau'\!-\!\tilde\beta'\!+\!\tilde\beta^*)S=\kappa'+(\tau^*\!-\!\beta^*\!+\!\beta')S+\del' S\;.
\end{align}
Before moving on, it is helpful to take stock of where we are. We have already solved directly for $\re(\tilde\e)$, $\im(\tilde\rho)$, and $\tilde \kappa$, and may thereby observe that they remain unshifted. We have also solved for $\re(\tilde\e')$ and $\im(\tilde\rho')$. By inserting the third equation in Eq.~(\ref{eq:second set of equations}) into the fourth one, we obtain the shifted $\kappa'$:
\begin{equation}
\tilde \kappa' = \kappa'+\left(\del'-2\beta^*\!+\!2\beta'\right)S+(\tau^*\!-\!\tau')S-\kappa^* S^2.
\end{equation}
Recall that $S \sim (-1,-1)$ and that the GHP-covariant version of $\del'$ is $\ETH' = \del'-2h\beta'+2h\beta^*$. As expected from GHP covariance, the NP derivatives and gauge fields appear in just the right combination to form a covariant derivative:
\begin{equation}\label{eq:shifted kappa' in GHP form}
\tilde \kappa' = \kappa' + \ETH'S+(\tau^*\!-\!\tau')S-\kappa^* S^2\;.
\end{equation}
On the other hand, the terms involving $D$, $\e$, and $\e^*$ in $\re(\tilde \e')$ do not collect themselves into a GHP-covariant combination. But that too is expected: While $\tilde\kappa'$ is a weighted quantity, $\tilde\e'$ is not. By solving the matrix inversion problem in Eq.~(\ref{eq:matrix inversion}) for the shifted frame, we obtain the shifted NP derivatives:
\begin{equation}
\tilde D = D\;,\;\; \tilde D' = D'- S\,D\;,\;\;\tilde\del = \del\;.
\end{equation}
We will see that $\tilde \e'$ will in fact combine with $\tilde D'$ to create a shifted $\tilde\THORN'$ that can be written in terms of GHP-covariant quantities. But to prove that, we will need to solve for the shifted $\im(\tilde\e')$, and for that we will need to study $dm$.
\\\\
Applying the above procedure to $dm$, we find the final set of shifted spin coefficient equations: 
%
%
%
\begin{align} \label{eq:third set of equations}
&\tilde\beta+\tilde\beta'^* = \beta+\beta'^*\;,\;\; \tilde\tau-\tilde\tau'^* = \tau-\tau'^*\;,\;\; \tilde\rho-2i\,\im(\tilde\e)=\rho-2i\,\im(\e)\;,\nonumber\\
&\tilde\rho'+2i\,\im(\tilde\e')+(\tilde\rho-2i\,\im(\tilde\e))S=\rho'+2i\,\im(\e')\;,\;\; \tilde\s'^*+\tilde\s S=\s'^*\;.
\end{align}
%
By solving Eqs.~(\ref{eq:first set of equations}),~(\ref{eq:second set of equations}), and~(\ref{eq:third set of equations}), we learn that the weighted spin coefficients and gauge field associated with $l_\mu$ do not receive corrections:
\begin{equation}\label{eq:shifted spin coefficients for l}
\tilde \kappa = \kappa\;,\;\; \tilde \tau = \tau\;,\;\; \tilde \s = \s\;,\;\; \tilde\rho = \rho\;,\;\; \tilde \e = \e\;.
\end{equation}
While it should not be surprising that $\kappa,\s,\rho$, and $\e$ do not receive corrections, it may be unexpected that $\tau$ does not shift. It turns out that $\tau'$ also remains unshifted:
\begin{equation}
\tilde\tau' = \tau'\;.
\end{equation}
So the timelike expansion $\tau+\tau'^*$ and the timelike twist $\tau-\tau'^*$ remain unshifted. 
\\\\
The weighted spin coefficients and gauge field associated with $l'_\mu$ do receive corrections:
\begin{align}
&\tilde \kappa' = \kappa'+\left[\ETH'+\left(\tau^*\!-\!\tau'\right)\right]S-\kappa^*S^2\;,\;\;\tilde\s' = \s'-\s^*S\;,\;\;\tilde \rho' = \rho' - \rho\, S\;,\nonumber\\
&\tilde\e' = \e'-\e^*S+\half DS-i\,\im(\rho)\,S\;.
\end{align}
In general, the transverse gauge fields also receive corrections:
\begin{equation}\label{eq:shifted beta}
\tilde\beta = \beta+\half \kappa\,S\;,\;\; \tilde\beta' = \beta'-\half\kappa^*S\;.
\end{equation}
From Eqs.~(\ref{eq:shifted spin coefficients for l})-(\ref{eq:shifted beta}) we conclude that if we align $l^\mu$ with background geodesics---namely if $\kappa = 0$---then not only do the formulas simplify considerably, but all nonlinearity in the shift function drops out of the spin coefficients. 
\\\\
This already implies $\tilde R_{abcd} = R_{abcd}+S\,R_{abcd}^{(1)}+ S^2\,R_{abcd}^{(2)}$, i.e., there are no terms of $O(S^3)$ or higher. Furthermore, if the geodesics to which $l^\mu$ are aligned can also be taken shear-free---namely if $\s = 0$---then we get $\tilde\s' = \s'$ as well. Finally, if we also align $l'^\mu$ with background shear-free geodesics, then
\begin{equation}\label{eq:shifted spin coefficients}
\tilde\kappa' = \left[\ETH'+\left(\tau^*\!-\!\tau'\right)\right]S\;,\;\; \tilde\rho'=\rho'-\rho\,S\;,\;\; \tilde\e'=\e'-\e^*S+\half DS-i\,\im(\rho)\,S\;.
\end{equation}
In this case, the only GHP-covariant derivative that shifts is $\THORN'$. The shifted version acting on a function $f_{h,\bar h} \sim (h,\bar h)$ is
\begin{equation}\label{eq:shifted thorn'}
\tilde\THORN' f_{h,\bar h} = \left[ \THORN' - S \THORN - (h\!+\!\bar h)(\THORN S)+2i(h\!-\!\bar h)\im(\rho) S\right] f_{h,\bar h}\;.
\end{equation}
This vindicates the discussion below Eq.~(\ref{eq:shifted kappa' in GHP form}) and completes our derivation of the shifted spin coefficients.
\\\\
Dray and 't~Hooft explained \cite{shockwave} that test particles crossing the shockwave get translated and refracted. (See also the work by Matzner \cite{matzner}.) In the spin coefficient formalism, these effects are described by the shifted versions of $\rho'$ and $\kappa'$---to the physics we now turn. 

\subsection{Shifted horizon}

Cartography of the horizon requires the hatted basis. As we discussed back in Sec.~\ref{sec:smooth tetrad}, the future horizon can be defined locally as the subspace of Kruskal-like coordinates on which the expansion of the outgoing congruence vanishes ($\hat \rho = 0$). Similarly, the past horizon is the subspace on which the expansion of the ingoing congruence vanishes ($\hat\rho' = 0$). 
\\\\
Recalling the unshifted $\hat\rho$ and $\hat\rho'$ from Eq.~(\ref{eq:hatted expansions}) and the shift described in Eq.~(\ref{eq:shifted spin coefficients}), we find that the coordinate $V$ receives a correction while the coordinate $U$ does not:
\begin{align}
&\tilde{\hat \rho} = \hat\rho \implies \tilde U = U\;,\nonumber\\
&\tilde{\hat{\rho}}' = \hat\rho'-\hat\rho\,\hat S = \left( 1+\frac{U}{V}\,\del(U)\,f(\ta,\chi)\right) \hat\rho' \equiv \left[ \frac{1}{2|R|^2}\left(\frac{\Delta}{UV}\right)\frac{1}{R^*}\right]\,\tilde V \nonumber\\
&\qquad\implies \tilde V-V=U\,\del(U)\,f(\ta,\chi)\;. \label{eq:diff eq for shifted V}
\end{align}
This last expression implies that smooth functions of $U$ will experience no coordinate shift, while functions that go as $\frac{1}{U}$ near $U = 0$ will experience a discontinuity in the coordinate. To see this, interpret Eq.~(\ref{eq:diff eq for shifted V}) as a differential equation in $U$ in the vicinity of $U = 0$, i.e., $\frac{d(\tilde V-V)}{dU} = \lim_{U\to 0}\frac{\tilde V-V}{U} = \del(U)\,f(\ta,\chi)$. Integration then gives
\begin{equation}\label{eq:shifted V}
\tilde V = V + \Theta(U)\,f(\ta,\chi)\;.
\end{equation}
This is the shift as described by Dray and 't Hooft \cite{shockwave} and by Sfetsos \cite{sfetsos}.

\subsection{Refraction}

Since every acolyte of Penrose knows that $\kappa$ and $\kappa'$ describe the refraction of light rays, the result that $\kappa'$ becomes nonzero after the shift speaks for itself. 

\section{Petrov classification for the Kerr-Newman shockwave}\label{sec:petrov}

Let the games begin. We will first shift the Weyl scalar $\Psi_4 \sim (-2,0)$, or more conveniently its complex conjugate $\Psi_4^* \sim (0,-2)$. Since this is just our opening act, we will reserve intricate computational details for the main event, the shifted Ricci scalars.

\subsection{Shifted $\Psi_4$ and physical interpretation}

Aligning the background frame with shear-free geodesic congruences but assuming an arbitrary shift function $S$, we find:
\begin{equation}\label{eq:shifted Psi4}
\tilde\Psi_4^* = \Psi_4^* + \ETH\ETH S + 2\tau\,\ETH S\;.
\end{equation}
To obtain this we used the complex conjugate and the prime of $\Phi_{02}$ from Eq.~(\ref{eq:Phi00}) in the forms $\ETH'\tau^* = -\tau^{*2}$ and $\ETH'\tau'=-\tau'^2$, which hold when $\Phi_{02} = 0$.
\\\\
To specialize to the shockwave, hat everything and insert the ansatz of Eq.~(\ref{eq:hatted shift ansatz}) for the shift function. Since the calculation is laborious, it is advantageous to first enumerate conceivable terms. 
\\\\
Remember that the horizon field $f(\ta,\chi)$ has weights $(-1,-1)$. Since $\Psi_4^*$ has weights $(0,-2)$, we will have to find operators of weights $(1,-1)$. Fortunately, the list of such operators that are nonzero at the Kerr-Newman horizon is short: 
\begin{equation}
\ETH\ETH\;;\;\; \tau \ETH\;,\;\; \tau'^*\ETH\;;\;\; \tau^2\;,\;\; \tau'^{*\,2}\;,\;\; \tau \tau'^*\;.
\end{equation}
In principle we would also need $\ETH \tau$ and $\ETH\tau'^*$, but again when $\Phi_{02} = 0$ those can be traded for $-\tau^2$ and $-\tau'^{*2}$. So the result must have the form
\begin{equation}\label{eq:general form of shifted Psi4*}
\hat{\tilde\Psi}_4^* = k_0\,\del(U)\left[ \ETH\ETH f+(k_1\, \tau + k_2\, \tau'^*)\ETH f +(k_3\, \tau^2 + k_4\, \tau'^{*\,2}+k_5\, \tau \tau'^*)f \right]
\end{equation}
for some functions $k_i(\ta)$ that will depend on the parameters $r_+$, $a$, and $\al$. Whether by hand or by machine we ultimately find:
\begin{align}\label{eq:shifted Psi4 for ansatz}
&k_0 = -\,\tfrac{c}{2|R|^2}\;,\;\;k_1 = \tfrac{\al |R|^2}{r}\;,\;\;k_2 = -2\left( 1+\tfrac{\al |R|^2}{2r}\right)\;,\;\; k_3 = \left( \tfrac{\al |R|^2}{2r}\right)^2\;,\nonumber\\
&k_4 = 1+\left( 1+\tfrac{\al |R|^2}{2r}\right)^2\;,\;\; k_5 = -\,\tfrac{\al |R|^2}{r}\left( 1+\tfrac{\al |R|^2}{2r}\right)\;.
\end{align}
On the way to this result, we encounter terms involving $\pa_U\del(U)$ and $\pa_U^{\,2}\del(U)$.\footnote{We also stumble upon the gargantuan notational implosion ``$\ETH\del(U) = \del \del(U)$.''} We interpret them according to the distributional edict of integrating by parts against an arbitrary smooth test function $\mathcal F(U)$:
\begin{equation}\label{eq:integrate against test function}
\int dU\, \mathcal F(U)\,\pa_U^{\;\;n} \del(U) = \int dU\,(-1)^n\,\pa_U^{\;\;n}\mathcal F(U)\,\del(U)\;.
\end{equation}
It should also be understood, as required by the overall factor $\del(U)$, that all instances of $r$ in Eq.~(\ref{eq:shifted Psi4 for ansatz}) actually denote $r_+$. Also note that numerically we have
\begin{equation}\label{eq:tau' vs tau*}
\tau' = -\,\frac{R}{R^*}\,\tau^*\;,
\end{equation}
so it is possible to shuffle terms among the coefficients $k_3, k_4$, and $k_5$. The particular form shown in Eq.~(\ref{eq:shifted Psi4 for ansatz}) is what we exhumed upon performing the rituals to be disclosed in Sec.~\ref{sec:sausage}.
\\\\
Invoking the gravitational compass from Sec.~\ref{sec:gravitational compass}, we interpret Eqs.~(\ref{eq:general form of shifted Psi4*}) and~(\ref{eq:shifted Psi4 for ansatz}) as describing a transverse ``outgoing'' gravitational wave stuck to the horizon.\footnote{We cannot help calling the reader's attention to the following famous quotation: ``Now, here, you see, it takes all the running you can do, to keep in the same place.'' This is originally from \textit{Through the Looking-Glass} by Lewis Carroll, but we first encountered its application to the horizon of a black hole from the textbook on the Kerr geometry by O'Neill \cite{o'neill}. 
}

\subsection{Nonrotating limit}

It is worth pausing to consider the nonrotating limit, $a \to 0$, in which case only the $\ETH\ETH f$ term in Eq.~(\ref{eq:general form of shifted Psi4*}) survives.
\\\\
As far as we know, the Weyl scalars for the shifted Reissner-Nordstr\"om geometry have not been calculated explicitly, so we will unpack the definitions of the GHP derivatives at the horizon. Remembering that $f \sim (-1,-1)$ and therefore $\ETH f \sim (-\half,-\tfrac{3}{2})$, and that in the nonrotating limit we have $\beta = \beta' = \beta^*=\beta'^*$, we find:
\begin{equation}
\left.\ETH\ETH f\right|_{a\,=\,0} = \del\del f-2\beta\,\del f\;.
\end{equation}
%

\subsection{Shifted $\Psi_3$ and Petrov type}

Our debt to $\Psi_4$ settled, we turn to $\Psi_3$. Shifting the frame (with $\kappa = \s = \kappa' = \s'$) seemingly produces this Weyl scalar:
\begin{align}\label{eq:shifted Psi3}
\tilde\Psi_3^* &= \Psi_3^*+\fourth(\THORN\ETH+\ETH\THORN)S+\fourth(2\tau\!-\!\tau'^*)\THORN S+\fourth(2\rho\!-\!5\rho^*)\ETH S -\half[(2\tau\!-\!\tau'^*)\rho^*+\ETH\rho^*]S\;.
\end{align}
But by hatting and specializing to Eq.~(\ref{eq:hatted shift ansatz}), we find that each term in Eq.~(\ref{eq:shifted Psi3}) goes to zero at $U = 0$ for fixed nonzero $V$:
\begin{equation}\label{eq:shifted Psi3 for ansatz}
\hat{\tilde\Psi}_3^* = 0\;.
\end{equation}
Since the unshifted geometry already had a nonzero $\Psi_2$, we conclude that the shockwave is Petrov type II:
\begin{equation}
\hat{\tilde\Psi}_0 = \hat{\tilde\Psi}_1 = \hat{\tilde\Psi}_3 = 0\;,\;\; \hat{\tilde\Psi}_2 \neq 0\;,\;\; \hat{\tilde\Psi}_4 \neq 0\;.
\end{equation}
To quote Szekeres: ``[I]t can be viewed as a Coulomb field with an outgoing wave component superimposed'' \cite{szekeres}.


%
%

\subsection{Curvatures of submanifolds}

Shifting both sides of the GHP commutator equations [see Eq.~(\ref{eq:formulas for K and Ks})], we find
\begin{equation}\label{eq:shifted K and Ks}
\tilde{\mathcal K} = \mathcal K+\im(\rho)\left[ 2\,\im(\rho)\,S+i\,(\THORN S)\right]\;,\;\;\tilde{\mathcal K_s} = \mathcal K_s-\half(\THORN^2 S)+i\,\THORN\left[ \im(\rho)S\right]\;.
\end{equation}
%
But if we hat everything and specialize to the shockwave ansatz, we will find that all of the corrections in Eq.~(\ref{eq:shifted K and Ks}) go to zero. Curiously enough, the shockwave does not alter the spacelike and timelike curvatures.

\subsection{Shifted $\Psi_2$}

Inserting Eq.~(\ref{eq:shifted K and Ks}) into Eq.~(\ref{eq:Psi2}) provides the shifted Weyl scalar of weight zero:
\begin{equation}
\tilde\Psi_2 = \Psi_2 + \sixth \THORN^2 S-\third(2\rho\!-\!\rho^*)\THORN S+\third\left[ (3\rho\!-\!2\rho^*)\rho+\Phi_{00}\right]S\;.
\end{equation}
To arrive at this expression, we used the relation\footnote{This is Raychaudhuri's equation for null shear-free geodesic congruences. When $\Phi_{00} = 0$, it tells us that $\THORN \rho = -\rho^2$. Given the standard interpretation of $\re(\rho)$ as the expansion, we recognize this as the focusing theorem.}
\begin{equation}\label{eq:Phi00 relation}
\Phi_{00} = -(\THORN\rho+\rho^2)\;\qquad(\text{if }\;\kappa = \s = 0)\;
\end{equation}
along with $\Phi_{00}^* = \Phi_{00}$. Just as we found for the shifted $\Psi_3$, we find upon disbursing hats and availing ourselves of Eq.~(\ref{eq:hatted shift ansatz}) that the correction to $\Psi_2$ is zero. 
\\\\
Appealing again to the gravitational compass \cite{szekeres}, we say that the Coulomb field remains unchanged by the presence of a massless particle on the future horizon.

\section{Shifted Ricci scalars}\label{sec:shifted ricci}

Show time. We will first present the shifted Ricci scalars for the generalized Kerr-Schild geometry under the assumption $\kappa = \kappa' = \s = \s'$, and then we will specialize to the shockwave.

\subsection{Ricci scalar of weight $(-1,-1)$: Absence of nonlinearity}

After the shift from Eq.~(\ref{eq:shifted tetrad}), three of the Ricci scalars will become nonzero. Of these, the apple of our eye will be $\Phi_{22} \sim (-1,-1)$. 
\\\\
This quantity is defined by priming the definition of $\Phi_{00}$ in Eq.~(\ref{eq:Phi00}):
\begin{equation}\label{eq:Phi22 GHP}
\Phi_{22} = -\left( \THORN'\rho'+\rho'^{\,2}\right)+\ETH\kappa'+\tau\kappa'+\tau' \kappa'^*-|\s'|^2\;.
\end{equation} 
Using the shifted $\rho'$ from Eq.~(\ref{eq:shifted spin coefficients}) and the shifted $\THORN'$ from Eq.~(\ref{eq:shifted thorn'}), and using $h = \bar h = -\half$ for $\rho'$ [recall Eq.~(\ref{eq:weighted spin coefficients for l'})], we find:
\begin{align}\label{eq:terms with rho'}
&\tilde\THORN'\tilde\rho'=\THORN'\rho'+(\rho'\THORN-\rho\THORN')S-(\THORN'\rho+\THORN\rho')S+(\THORN\rho)S^2\;,\nonumber\\
&\tilde\rho'^2 = \rho'^2-2\rho\rho' S + \rho^2 S^2\;.
\end{align}
It is worth keeping in mind the formula for $\Phi_{00}$ under the shear-free geodesic assumption [Eq.~(\ref{eq:Phi00 relation})]. Next, for $\s\s' = \kappa\kappa' = 0$, we have:\footnote{Attentive readers have every right to be confused by the second equality: Indeed it turns out that the combination of derivatives and products of spin coefficients in Eq.~(\ref{eq:(4.12.32)(f)}) equals its primed version in Eq.~(\ref{eq:(4.12.32)(f')}). This must be so, since both $\Psi_2 = C_{1342}$ and $\Pi = \tfrac{1}{12}(R_{12}-R_{34})$ are self-prime.}
\begin{align}
\Psi_2+2\Pi &= -(\ETH'\tau+|\tau|^2)+\THORN'\rho+\rho'^*\rho \label{eq:(4.12.32)(f)}\\
&= -(\ETH\tau'+|\tau'|^2)+\THORN\rho'+\rho^*\rho'\;. \label{eq:(4.12.32)(f')}
\end{align}
%
It is a matter of some discretion which variables to keep and which to trade away. We are guided by comparison with the nonrotating limit, which suggests we should express as much as possible in terms of $\tau$ and $\tau'$ and their derivatives. So we will use Eqs.~(\ref{eq:(4.12.32)(f)}) and~(\ref{eq:(4.12.32)(f')}) to evict $\THORN'\rho$ and $\THORN\rho'$ from Eq.~(\ref{eq:terms with rho'}). 
\\\\
With our shifted $\kappa'$ from Eq.~(\ref{eq:shifted spin coefficients}), we find:
\begin{align}\label{eq:terms with kappa'}
&\ETH\tilde\kappa' = \ETH\ETH'S+(\tau^*\!-\!\tau')\ETH S+(\ETH\tau^*\!-\ETH\tau')S\;,\nonumber\\
&\tau\tilde\kappa'+\tau'\tilde\kappa'^*=(\tau\ETH'+\tau'\ETH)S+(|\tau|^2\!-\!|\tau'|^2)S\;.
\end{align}
Eqs.~(\ref{eq:terms with rho'})-(\ref{eq:terms with kappa'}) then supply the preliminary expression:
\begin{align}\label{eq:shifted Phi22 preliminary}
\tilde\Phi_{22} &= \Phi_{22} + (\rho\THORN'-\rho'\THORN)S+\Phi_{00}S^2+\ETH\ETH' S + \tau \ETH' S + \tau^* \ETH S\;\nonumber\\
&+\left[ 2\rho\rho'-(\rho\rho'^*\!+\!\rho^*\rho')+\ETH'\tau+\ETH\tau^*+2|\tau|^2+2(\Psi_2+2\Pi)\right] S\;.
\end{align}
Behold: For a background in which $\Phi_{00} = 0$, \textit{all nonlinear dependence on the perturbation drops out of the curvature scalars}. Terms of $O(S^2)$ could not possibly show up elsewhere, because the only curvature scalar with the appropriate weight to include a \textit{product} of shifted quantities (in this case $\THORN'$ and $\rho'$) is $\Phi_{22}$. 
\\\\
To make sense of Eq.~(\ref{eq:shifted Phi22 preliminary}) we will rewrite it in a manifestly real form:
\begin{align}\label{eq:shifted Phi22}
\tilde\Phi_{22} = \re(\tilde\Phi_{22}) &= \Phi_{22}+\re\left[(\rho\THORN'-\rho'\THORN)S\right]+\Phi_{00} S^2 \nonumber\\
&+ \half\left[ \ETH\ETH'+\ETH'\ETH+(\tau\!+\!\tau'^*)\ETH'+(\tau^*\!+\!\tau')\ETH + (\tau\!-\!\tau'^*)\ETH'+(\tau^*\!-\!\tau')\ETH\right] S \nonumber\\
&+\left\{ (\rho\!-\!\rho^*)(\rho'\!-\!\rho'^*)+(\ETH'\tau+c.c.) + 2|\tau|^2+2\left[\re(\Psi_2)+2\Pi\right]\right\} S\;.
\end{align}
Experts in the compacted formalism should recognize the combination $\ETH\ETH'+(\tau\!+\!\tau'^*)\ETH'+c.c.$ as part of the generalized Laplacian (we will get to this in Sec.~\ref{sec:sausage}). Before elaborating on this, we will vanquish the remaining curvature scalars.

\subsection{Other Ricci scalars}

The Ricci scalar of weight $(-1,0)$ is corrected by the general shift:\footnote{The steps leading to this expression parallel closely those that led to $\tilde\Psi_3^*$.}
\begin{align}\label{eq:shifted Phi21}
\tilde\Phi_{21} &= \Phi_{21}+\fourth(\THORN\ETH'+\ETH'\THORN)S+\fourth(2\tau^*\!-\!\tau')\THORN S + \fourth(3\rho\!-\!2\rho^*)\ETH'S+\half(\tau'\rho\!-\!2\tau^*\rho^*\!+\!\ETH'\rho)S\;.
\end{align}
For the Kerr-Newman background, we have $\Phi_{21}= 0$. After hatting and specializing to Eq.~(\ref{eq:hatted shift ansatz}), we find that each would-be contribution from $S$ to Eq.~(\ref{eq:shifted Phi21}) is zero. 
\\\\
Next we have the Ricci scalar of weight $(0,0)$:
\begin{equation}
\tilde\Phi_{11} = \Phi_{11} + \fourth\THORN^2 S - \half\left[ |\rho|^2+(\rho\!-\!\rho^*)^2\right]S\;.
\end{equation}
Here too we find no correction to the unshifted value after hatting both sides of the equation and specializing to the shockwave: $\hat{\tilde\Phi}_{11} = \hat \Phi_{11} = \Phi_{11}$.
\\\\
The Einstein-Hilbert curvature also superficially becomes nonzero as a result of the shift:
\begin{equation}\label{eq:shifted Pi}
\tilde\Pi = \Pi-\sixth\left\{ \half \THORN^2 S+(\rho\!+\!\rho^*)\THORN S +(|\rho|^2-2\Phi_{00})\,S \right\}\;.
\end{equation}
But we know that $\Pi$ is proportional to the Lagrangian of general relativity, so its first order variation must comport with the standard formula
\begin{equation}\label{eq:vary the action}
S_{\text{GR}}[g+h]- S_{\text{GR}}[g] = \half\int \!d^4x\,|\det(g)|^{1/2}\, T^{\mu\nu} h_{\mu\nu} + O(h^2)\;. 
\end{equation}
The shift from Eq.~(\ref{eq:shifted tetrad}) effects the metric variation
\begin{equation}
h_{\mu\nu} = -2S\, l_\mu l_\nu\;.
\end{equation}
So varying the action with respect to $S$ will result in something proportional to $T^{\mu\nu} l_\mu l_\nu = \mathcal T^{\mu\nu} l_\mu l_\nu = (8\pi)^{-1}\, t_{00}$ (recall Sec.~\ref{sec:energy scalars}). 
Because the only nonzero energy scalar for the background spacetime is $t_{11} \propto (l_\mu l'_\nu+l'_\mu l_\nu + m_\mu m'_\nu + m'_\mu m_\nu )\mathcal T^{\mu\nu}$, we know that $t_{00} = 0$ and thereby expect the $O(S)$ term in Eq.~(\ref{eq:vary the action}) to equal zero.\footnote{We thank Alexei Kitaev for suggesting this check on our work.}
\\\\
The nonzero $O(S)$ term in Eq.~(\ref{eq:shifted Pi}) might invite consternation, but we have been cavalierly ignoring possible boundary terms in the action. So all we require is that the $O(h)$ term in Eq.~(\ref{eq:vary the action}) should be zero, not necessarily that the shift in $\Pi$ itself should be zero. 
\\\\
For Kerr-Newman, we have\footnote{This expression for $|\det(g)|^{1/2}$ makes clear that it does not receive a correction from Eq.~(\ref{eq:shifted tetrad}).} $|\det(g)|^{1/2} = i\,\e^{\mu\nu\rho\s} l_\mu l'_\nu m_\rho m'_\s = |R|^2\sin\ta$. After integrating by parts, dropping total derivatives, and using $D$ and $\rho$ from Eqs.~(\ref{eq:vectors}) and~(\ref{eq:spin coefficients for kerr-newman}), we indeed obtain 
\begin{equation}
\int d^4x\,|\det(g)|^{1/2}\,\tilde \Pi = 0\;.
\end{equation}
This completes our account of the shifted curvature scalars for the generalized Kerr-Schild geometry. (The Ricci scalars not explicitly enumerated in this section do not shift.) Now we will specialize the shifted $\Phi_{22}$ to the shockwave.

\section{Derivatives of the shift}\label{sec:sausage}

The spacetime Laplacian $\na^2 = \na_\mu \na^\mu$ finds refuge in the compacted spin coefficient formalism within a more general operator
\begin{equation}\label{eq:parts of box}
\square = -\square_\parallel + \square_\perp\;,
\end{equation}
where
\begin{equation}\label{eq:box decomposition}
\square_\parallel \equiv [\THORN+2\,\re(\rho)]\THORN'\;+\;'\;,\;\;\square_\perp \equiv [\ETH+(\tau+\tau'^*)]\ETH'\;+\;'\;.
\end{equation}
The operator $\square_\perp$ will be called the ``transverse box.'' Evaluating its action on the shift function is the most technically cumbersome aspect of computing $\tilde \Phi_{22}$.
\\\\
We will do our best to show how the sausage is made without belaboring mindless \-algebra.

\subsection{Key facts}

To set up the calculation we will first collect some useful formulas. 
\\\\
From what may seem like a lifetime ago, we recall that $U\pa_U r = V\pa_V r$ (which can be traced back to the relation $-U\pa_U + V\pa_V = \frac{1}{\al}\pa_t$). Therefore, acting on a weight-$(0,0)$ function $F(r)$, we have:
\begin{equation}\label{eq:key fact 1}
\ETH F(r) = 0\;.
\end{equation}
This is our first key fact.
\\\\
Next we recall the explicit formulas for the timelike expansion and the timelike twist [Eq.~(\ref{eq:tau and tau'})]. They will compose our basic mnemonic for making sense of complicated algebraic expressions: The trigonometric functions $\sin(2\ta)$ and $\sin\ta$ should evoke $\tau\!+\!\tau'^*$ and $\tau\!-\!\tau'^*$ respectively. 
\\\\
We will use this to establish additional useful formulas. Treating $\del(U)$ as having weight $(0,0)$ and summoning the NP derivatives in Kruskal-like coordinates [Eq.~(\ref{eq:kruskal vectors})], we find:
\begin{equation} \label{eq:key fact 2}
\ETH \del(U) = -\,\al\,\frac{i a \sin\ta}{R\sqrt 2}\,U\pa_U\del(U) = -\,\frac{\al |R|^2}{2r}\,(\tau\!-\!\tau'^*)\,U\pa_U\del(U)\;.
\end{equation}
This is our second key fact. 
\\\\
Finally, we must bear in mind that although functions of $r$ can be treated as constants, the generalized radial function $R = r+ia\cos\ta$ is also a function of $\ta$. Treating this too as a function of weight $(0,0)$, we compute the following:
\begin{equation}\label{eq:key fact 3}
\ETH\left( \frac{1}{|R|^2}\right) = -\,\frac{1}{(|R|^2)^2}\,\ETH\!\left(|R|^2\right) = +\,\frac{a^2\sin(2\ta)}{\sqrt 2 R|R|^4} = -\,\frac{1}{|R|^2}(\tau\!+\!\tau'^*)\;.
\end{equation}
This is our third key fact.

\subsection{Integration by parts}

We described back in Eq.~(\ref{eq:integrate against test function}) the standard integration-by-parts procedure that defines the delta function. Here it will be useful to study two special cases of that formula.
\\\\
First consider a distribution $\op(U)\, U\pa_U\del(U)$ (where the conditions on $\op(U)$ will be specified shortly), and integrate it against a test function $\mathcal F(U)$ that falls off quickly enough to merit dropping the boundary term:
\begin{equation}
\int dU\,\op(U)\,U\pa_U \del(U)\,\mathcal F(U) = -\int dU\,\left[ \op(U)\mathcal F(U)+U\pa_U(\op(U) \mathcal F(U))\right]\del(U)\;.
\end{equation}
If $\pa_U(\op(U)\mathcal F(U)) \sim U^n$ with $n \geq 0$ near $U = 0$, then the second term evaluates to zero. We then obtain the following distributional equality:
\begin{equation}\label{eq:distributional equality 1}
\op(U)\,U\pa_U \del(U) = -\op(U)\,\del(U)\;.
\end{equation}
Along similar lines, we will obtain a second distributional equality:
\begin{equation}\label{eq:distributional equality 2}
\op(U)\,U\pa_U(U\pa_U \del(U)) = +\op(U)\,\del(U)\;.
\end{equation}
Equipped with the key facts in Eqs.~(\ref{eq:key fact 1})-(\ref{eq:key fact 3}) and the above distributional equalities, we are ready to face the transverse box.

\subsection{First-derivative terms}

We warm up with a first-derivative term. Specializing to the shockwave ansatz in Eq.~(\ref{eq:hatted shift ansatz}) and applying our key facts, we obtain the preliminary expression 
\begin{align}\label{eq:eth S}
\ETH \hat S &= \half \tfrac{\Del}{UV}\left[ \ETH(\tfrac{1}{|R|^2})\del(U) f(\ta,\chi)+\tfrac{1}{|R|^2}\ETH\del(U) f(\ta,\chi)+\tfrac{1}{|R|^2}\del(U)\ETH f(\ta,\chi)\right] \nonumber\\
&= \tfrac{1}{2|R|^2}\tfrac{\Del}{UV}\left\{\del(U)\left[ \ETH-(\tau\!+\!\tau'^*)\right]f(\ta,\chi)-\tfrac{\al |R|^2}{2r}(\tau\!-\!\tau'^*)\,U\pa_U \del(U)\,f(\ta,\chi)\right\}\;.
\end{align}
Before integrating by parts against a test function, we need to multiply by $\tau^*+\tau'$ to obtain the term $(\tau^*\!+\!\tau')\ETH\hat S$ that appears in the transverse box.\footnote{Since $\tau^*\!+\tau'$ depends on $U$ and $V$ only through $r = r(UV)$, and since we have already said such functions can be treated as constants with respect to $U\pa_U$ for our calculation, it does not matter in this particular instance whether we integrate by parts before or after multiplying by $\tau^*\!+\tau'$.}
\\\\
Note that since $|\tau|^2 = |\tau'|^2$ for the Kerr-Newman spacetime, we have
\begin{equation}
(\tau^*\!+\!\tau')(\tau\!-\!\tau'^*) = 2i\,\im(\tau\tau')\;.
\end{equation}
Using this and the distributional equality in Eq.~(\ref{eq:distributional equality 1}), we obtain [also recall $c \equiv -\left.\frac{\Del}{UV}\right|_{r\,=\,r_+}$ from Eq.~(\ref{eq:Delta/UV})]
\begin{equation}\label{eq:(tau*+tau')eth S}
(\tau^*\!+\!\tau')\,\ETH \hat S = -\,\frac{c}{2|R|^2}\,\del(U)\left\{ (\tau^*\!+\!\tau')\,\ETH-|\tau\!+\!\tau'^*|^2+i\,\tfrac{\al |R|^2}{r}\,\im(\tau\tau') \right\}f(\ta,\chi)\;.
\end{equation}

\subsection{Second-derivative terms}

Returning to Eq.~(\ref{eq:eth S}), we act with $\ETH'$ (and skip a few steps now that the method is presumably clear) to obtain
\begin{align}\label{eq:eth'eth S}
\ETH'\ETH \hat S &= \tfrac{1}{2|R|^2}\tfrac{\Del}{UV}\left\{ \del (U)\,\ETH'\ETH f - \left[(\tau\!+\!\tau'^*)\,\del(U)+\tfrac{\al |R|^2}{2r}(\tau\!-\!\tau'^*)\,U\pa_U\del(U) \right] \ETH' f \right. \nonumber\\
&\left. -\left[ (\tau^*\!+\!\tau')\,\del(U)+\tfrac{\al|R|^2}{2r}(\tau^*\!-\!\tau')\,U\pa_U\del(U)\right] \ETH f + \Cs\,f\right\}\;,
\end{align}
where
\begin{align}
\Cs &=|\tau\!+\!\tau'^*|^2 \,\del(U)-i\tfrac{\al |R|^2}{r}\im(\tau\tau')\,U\pa_U\del(U)-\left(\del(U)+\tfrac{\al |R|^2}{2r}\,U\pa_U\del(U)\right)\ETH'\tau \nonumber\\
&-\left(\del(U)-\tfrac{\al |R|^2}{2r}\,U\pa_U\del(U)\right)(\ETH \tau')^*+\left( \tfrac{\al |R|^2}{2r}\right)^2|\tau\!-\!\tau'^*|^2\, U\pa_U\left[U\pa_U\del(U)\right] \;.
\end{align}
Note that in Eq.~(\ref{eq:eth'eth S}) the coefficient of $\ETH' f$ is the complex conjugate of the coefficient of $\ETH f$. This did not have to be so, because we are computing $\ETH'\ETH\hat S$ right now, not $\ETH'\ETH\hat S + c.c.$, and in general $\ETH'\ETH \neq \ETH \ETH'$. 
\\\\
This quantity $\ETH'\ETH\hat S$ will be integrated directly against a test function (because it appears directly in the transverse box, which in turn appears directly in $\hat{\tilde\Phi}_{22}$), so we can use the distributional equalities in Eqs.~(\ref{eq:distributional equality 1}) and~(\ref{eq:distributional equality 2}), loosely expressed as $U\pa_U\del(U) \to -\del(U)$ and $U\pa_U[U\pa_U\del(U)] \to +\del(U)$. Applying these to Eq.~(\ref{eq:eth'eth S}), we obtain:
\begin{align}
&\ETH'\ETH\hat S = -\,\frac{c}{2|R|^2}\,\del(U)\left\{  \ETH'\ETH -\left[ (\tau\!+\!\tau'^*)-\tfrac{\al |R|^2}{2r}(\tau\!-\!\tau'^*)\right]\ETH' -\left[ (\tau^*\!+\!\tau')-\tfrac{\al |R|^2}{2r}(\tau^*\!-\!\tau')\right]\ETH  \right. \nonumber\\
& \left. +|\tau\!+\!\tau'^*|^2\!+i\tfrac{\al |R|^2}{r}\im(\tau\tau')-\left( 1\!-\!\tfrac{\al|R|^2}{2r}\right)\ETH'\tau-\left( 1\!+\!\tfrac{\al |R|^2}{2r}\right)(\ETH\tau')^*+\left( \tfrac{\al |R|^2}{2r}\right)^2|\tau\!-\!\tau'^*|^2\right\}\! f(\ta,\chi)\,.
\end{align}

\subsection{Transverse box}

Now we can finish the job. Returning to the first-derivative term in Eq.~(\ref{eq:eth S}) and adding its complex conjugate, we obtain:
\begin{equation}\label{eq:transverse box part 1}
(\tau^*\!+\!\tau')\ETH\hat S + c.c. = -\,\frac{c}{2|R|^2}\,\del(U)\left\{ (\tau\!+\!\tau'^*)\ETH'+(\tau^*\!+\!\tau')\ETH-2|\tau\!+\!\tau'^*|^2 \right\} f(\ta,\chi)\;.
\end{equation}
Next we obtain the anticommutator of GHP derivatives by taking Eq.~(\ref{eq:eth'eth S}) plus its complex conjugate:
\begin{align}\label{eq:transverse box part 2}
&\ETH'\ETH \hat S + c.c. = -\,\frac{c}{2|R|^2}\,\del(U)\left\{  (\ETH'\ETH+\ETH\ETH')+\left[  -\left(2(\tau\!+\!\tau'^*)-\tfrac{\al |R|^2}{r}(\tau\!-\!\tau'^*) \right) \ETH'+c.c.\right]\right. \nonumber\\
&\left. +2|\tau\!+\!\tau'^*|^2-\left( 1\!-\!\tfrac{\al |R|^2}{2r}\right)(\ETH'\tau + c.c.)-\left(1\!+\!\tfrac{\al |R|^2}{2r}\right)(\ETH\tau'+c.c.)+2\left( \tfrac{\al |R|^2}{2r}\right)^2 |\tau\!-\!\tau'^*|^2\right\} f(\ta,\chi)\,.
\end{align}
We then add Eqs.~(\ref{eq:transverse box part 1}) and~(\ref{eq:transverse box part 2}) to obtain the transverse box. For reasons morally unbeknownst to us, the $|\tau+\tau'^*|^2$ term will cancel out. Also, for Kerr-Newman, we have
\begin{equation}\label{eq:eth tau' = eth' tau}
\ETH\tau' = \ETH'\tau\;.
\end{equation}
There is probably a good reason for this, but it escapes us. At any rate, it implies that the $\al$-dependent parts of the coefficients of $\ETH'\tau+c.c.$ and $\ETH\tau'+c.c.$ drop out. 
\\\\
Therefore, the transverse box acting on the shift function, expressed in terms of GHP derivatives at the horizon, simplifies to:
\begin{align}\label{eq:transverse box S}
\square_\perp \hat S &= -\,\frac{c}{2|R|^2}\,\del(U)\left\{  (\ETH'\ETH+\ETH\ETH')+\left[  -\left((\tau\!+\!\tau'^*)-\tfrac{\al |R|^2}{r}(\tau\!-\!\tau'^*) \right) \ETH'+c.c.\right]\right. \nonumber\\
&\left. -2(\ETH'\tau+c.c.) +2\left(\tfrac{\al |R|^2}{2r} \right)^2|\tau\!-\!\tau'^*|^2\right\} f(\ta,\chi)\;.
\end{align}
This completes the most arduous part of the calculation. It bears repeating that all quantities in Eq.~(\ref{eq:transverse box S}) are understood to be evaluated at $r = r_+$, as mandated by the overall delta function.

\subsection{Laplacian on the squashed sphere}

We could leave the result for $\square_\perp \hat S$ in the form of Eq.~(\ref{eq:transverse box S}), but those familiar with the Dray-'t Hooft solution expect 2d Laplacians.
\\\\
Our shift function $S$ and our horizon field $f$ have GHP weight $(-1,-1)$. In general, a weighted function $f_{h,h} \sim (h,h)$ has \textit{spin-weight} $s \equiv h-\bar h = h-h = 0$. The shockwave has $h = -1$, but without much fuss we can understand the situation for $s = 0$ but arbitrary $h$.\footnote{Since complex conjugation exchanges $h$ and $\bar h$, only functions with $s = 0$ can be taken real. We therefore assume $f_{h,h}^* = f_{h,h}$ for simplicity.}
\\\\
By explicit computation on a function $f_{h,h}(\ta,\chi)$ of the Kruskal-like angular coordinates only, we find that the following combination of NP derivatives and GHP gauge fields reproduces the Laplacian on the \textit{squashed sphere}:\footnote{A squashed sphere of radius $r$ has line element $ds^2 = |R|^2 d\ta^2+\frac{|R_0|^4}{|R|^2} \sin^2\ta\, d\chi^2 \equiv h_{ij}\,dx^i\,dx^j$, and the Laplacian derived from that is $\na_{\text{2d}}^{\;2} = \frac{1}{|R|^2}\left[ \pa_\ta^{\,2}+\left( \frac{|R_0|^2+a^2\sin^2\ta}{|R|^2}\right)\cot\ta\,\pa_\ta+\frac{|R|^4}{|R_0|^4}\,\frac{1}{\sin^2\ta}\,\pa_\chi^{\,2}\right]$. If $R_{ij}^{\text{2d}}$ is the Ricci tensor derived from $h_{ij}$, then $\fourth h^{ij} R_{ij}^{\text{2d}} = \left.\re(\mathcal K) \right|_{r\,=\,r_+}$, the intrinsic curvature from Eq.~(\ref{eq:intrinsic curvature at horizon}).
%
%
}
\begin{equation}
\del'\del + \del\del'-(\beta+\beta'^*)\del'-(\beta'+\beta^*)\del = \na_{\text{2d}}^{\;2}\;.
\end{equation}
So unpacking the GHP derivatives according to their original definitions back in Eq.~(\ref{eq:GHP covariant derivatives}) provides the desired expression:
\begin{align}\label{eq:anticommutator in terms of 2d laplacian}
(\ETH'\ETH+\ETH\ETH') &f_{h,h}(\ta,\chi) = \left\{  \na_{\text{2d}}^{\;2}+\left[ 4h(\beta-\beta'^*) \del' + c.c.\right]\right. \nonumber\\
&\left. + 2h\left[ (\del'\beta-\del\beta' + c.c.)+2(|\beta'|^2\!-\!|\beta|^2)+4h|\beta\!-\!\beta'^*|^2\right] \right\} f_{h,h}(\ta,\chi)\;.
\end{align}
That is how our coveted 2d spatial Laplacian manifests in our story. Its tragedy is that while we may find temporary solace in a familiar face, this yearning for camaraderie cost us the guidance of GHP covariance, without which we are hopelessly lost. 
\section{Ricci tensor}\label{sec:ricci tensor}

The trace-reversed Ricci tensor, being necessary to the gravitational field of a localized Source, the propensity of a massless particle to generate Curvature, shall now be realized.

\subsection{Relation to curvature scalars}

We emerge from the chrysalis of the internal space by translating the usual prescription $R_{\mu\nu} = e_\mu^{\;a} e_\nu^{\;b} R_{ab}$ into NP notation:
\begin{align}\label{eq:ricci tensor in terms of ricci scalars}
\half R_{\mu\nu} &= l'_\mu l'_\nu\, \Phi_{00} + l_\mu l_\nu\, \Phi_{22}+\left[ m'_\mu m'_\nu\,\Phi_{02}-(l'_\mu m'_\nu\!+\!m'_\mu l'_\nu)\,\Phi_{01}-(l_\mu m_\nu\!+\!m_\mu l_\nu)\,\Phi_{21}+c.c.\right] \nonumber\\
&+(l_\mu l'_\nu\!+\!l'_\mu l_\nu\!+\!m_\mu m'_\nu \!+\! m'_\mu m_\nu)\,\Phi_{11}+(l_\mu l'_\nu\!+\!l'_\mu l_\nu\!-\!m_\mu m'_\nu\!-\!m'_\mu m_\nu)\,3\Pi\;.
\end{align}
To evaluate the right-hand side, we first need to tilde everything (to calculate shifted quantities), and then we need to hat everything (to work in the horizon basis). 
\\\\
We will specialize directly to the shockwave, so the only Ricci scalar that will shift is $\Phi_{22}$. Meanwhile, the unshifted geometry has only a nonzero $\Phi_{11}$. Therefore, we have for the full (i.e., including the unshifted part) Ricci tensor:
\begin{align}
\half {\tilde R}_{\mu\nu} &= \hat{\tilde l}_\mu \hat{\tilde l}_\nu \,\hat{\tilde\Phi}_{22}+(\hat{\tilde l}_\mu \hat{\tilde l}'_\nu + \hat{\tilde l}'_\mu \hat{\tilde l}_\nu + \hat{\tilde m}_\mu \hat{\tilde m}'_\nu + \hat{\tilde m}'_\mu \hat{\tilde m}_\nu) \,\hat{\tilde\Phi}_{11} \nonumber\\
&= \hat l_\mu \hat l_\nu\, \hat{\tilde\Phi}_{22} + (\hat l_\mu \hat{\tilde l}'_\nu+\hat{\tilde l}'_\mu \hat l_\nu + m_\mu m'_\nu +  m'_\mu  m_\nu)\, \Phi_{11}\;.
\end{align} 
In the second line we have removed the tildes for quantities that equal their unshifted counterparts, and we have removed the hats from quantities that do not get rescaled by factors of $U$ when passing from the standard frame to the horizon one.\footnote{There is no need to place a hat on the Ricci \textit{tensor}, because by construction it is invariant under GHP transformations of the frame.}
\\\\
Recalling from Eq.~(\ref{eq:shifted tetrad}) the premise that launched this travail in the first place, we isolate the part of the Ricci tensor that results from the shift:
\begin{equation}\label{eq:Ricci tensor from the shift}
R_{\mu\nu}^{\text{ shift}} = 2\,\hat l_\mu \hat l_\nu \,(\hat{\tilde\Phi}_{22} + 2\hat S\, \Phi_{11})\;.
\end{equation}
Returning to our explicit expressions for the 1-forms in Eq.~(\ref{eq:kruskal forms}), we find:
\begin{equation}
\left.\hat l \right|_{U\,=\,0} = \frac{|R_+|^2}{\al |R_{0+}|^2}\,dU\;.
\end{equation}
So we learn first of all that $R_{\mu\nu}^{\text{ shift}} = R_{UU}^{\text{ shift}}\,\del_\mu^{\;U} \del_\nu^{\;U}$, as promised.

\subsection{Relation to energy scalars}

Meanwhile, the energy tensor also admits an expansion analogous to Eq.~(\ref{eq:ricci tensor in terms of ricci scalars}):
\begin{align}
4\pi T_{\mu\nu} &= l'_\mu l'_\nu \,t_{00}+l_\mu l_\nu\, t_{22}+\left[ m'_\mu m'_\nu \,t_{02}-(l'_\mu m'_\nu\!+\!m'_\mu l'_\nu)\,t_{01}-(l_\mu m_\nu\! +\! m_\mu l_\nu)\,t_{21}+c.c.\right] \nonumber\\
&+(l_\mu l'_\nu\!+\! l'_\mu l_\nu \!+\! m_\mu m'_\nu \!+\! m'_\mu m_\nu)\, t_{11}+(l_\mu l'_\nu\!+\!l'_\mu l_\nu\!-\!m_\mu m'_\nu\!-\!m'_\mu m_\nu)\,3t_\Pi\;.
\end{align}
Anticipating the required energy tensor term by term, we conclude:
\begin{equation}
8\pi T^{\text{ shift}}_{\mu\nu} = 2\,\hat l_\mu \hat l_\nu\,(\hat{\tilde t}_{22}+2\hat S \,t_{11})\;.
\end{equation}
Given that the background Einstein equation is, by construction, $\Phi_{11} = t_{11}$ [recall Eq.~(\ref{eq:einstein eq})], all we need is a $t_{22}$ such that 
\begin{equation}\label{eq:Phi22 = t22}
\hat{\tilde \Phi}_{22} = \hat{\tilde t}_{22}\;.
\end{equation}
The whole point of this tale is that the correction to the left-hand side can be interpreted as the backreaction from a \textit{massless particle on the future horizon}, so that is what will populate the right-hand side. 
In this paper we focus on the geometry instead of the field theory, so let us leave that aside and press on. 

\subsection{Final result for $\Phi_{22}$}

Returning to our earlier calculation of $\ETH\hat S$ [Eq.~(\ref{eq:eth S})], multiplying by $\tau^*\!-\!\tau'$, integrating by parts, and adding the complex conjugate, we obtain the remaining first-derivative terms:
\begin{align}\label{eq:(tau-tau'*)eth'S}
(\tau\!-\!\tau'^*)\ETH'\hat S + c.c. &= -\,\frac{c}{2|R|^2}\,\del(U)\left\{ (\tau\!-\!\tau'^*)\ETH'+(\tau^*\!-\!\tau')\ETH + \tfrac{\al |R|^2}{r}|\tau\!-\!\tau'^*|^2 \right\} f(\ta,\chi)\;.
\end{align}
Next take the general shifted $\Phi_{22}$ from Eq.~(\ref{eq:shifted Phi22}), hat it, and recognize that $\hat\rho'\,\hat\THORN \hat S$ and $(\hat\rho\!-\!\hat\rho^*)(\hat\rho'\!-\!\hat\rho'^*)$ go to zero at $U = 0$. 
\\\\
But $\hat\rho\,\hat\THORN'\hat S$ is more subtle, since within $\hat D'$ lurks $\pa_U$. Applying Eq.~(\ref{eq:distributional equality 1}), we obtain
\begin{align}\label{eq:missing term}
\re(\hat\rho \hat D'\hat S) &= \al r_+\frac{|R_{0+}|^2}{|R_+|^4}\hat S = -\half\left.\left[\ETH\tau+\ETH'\tau^*+2|\tau|^2+2\,\re(\Psi_2)\right]\right|_{r\,=\,r_+}\hat S\;.
\end{align} 
Because $\left.\hat\rho\right|_{U\,=\,0} = 0$, the terms involving $\hat \e'$ and $\hat\e'^*$ drop out, leaving us with $\re(\hat\rho \hat\THORN'\hat S) = \re(\hat\rho \hat D'\hat S) = -\re(\hat\THORN'\hat\rho)\hat S$. 
\\\\
Putting all this together (and using $\Phi_{00} = \Pi = 0$), we reduce our shifted $\Phi_{22}$ to the relatively compact form:
\begin{align}
\hat{\tilde \Phi}_{22} &= \half\left\{\square_\perp + (\tau\!-\!\tau'^*)\ETH'+(\tau^*\!-\!\tau')\ETH +(\ETH'\tau+c.c.)+2|\tau|^2+2\,\re(\Psi_2)\right\}\hat S\;.
\end{align}
Enlisting our result for $\square_\perp\hat S$ in Eq.~(\ref{eq:transverse box S}) and the relation $4|\tau|^2 = |\tau\!+\!\tau'^*|^2+|\tau\!-\!\tau'^*|^2$, we finally obtain the beautiful, exquisite, magical expression
\begin{align}\label{eq:beautiful Phi22}
\hat{\tilde\Phi}_{22} &= -\,\frac{c}{4|R|^2}\,\del(U)\,\Ds f(\ta,\chi)\;,
\end{align}
where the differential operator $\Ds$ is
\begin{align}\label{eq:result}
\Ds &= \ETH'\ETH+\ETH\ETH'+ \left[-(\tau\!+\!\tau'^*)+\left(1+\tfrac{\al |R|^2}{r}\right)(\tau\!-\!\tau'^*) \right]\ETH'+\left[-(\tau^*\!+\!\tau')+\left(1+\tfrac{\al |R|^2}{r}\right)(\tau^*\!-\!\tau') \right]\ETH \nonumber\\
&+2\,\re(\Psi_2) - (\ETH'\tau+\ETH\tau^*)+\half |\tau\!+\!\tau'^*|^2+ \half\left(1+\tfrac{\al |R|^2}{r}\right)^2|\tau\!-\!\tau'^*|^2\;.
\end{align}
This is our final result. 
\\\\
It is expressed in terms of quantities that have innate geometrical significance, in that each operator has a definite GHP weight. When $a = 0$, we obtain\footnote{At $r = r_+$, we have $\left.\re(\Psi_2)\right|_{a\,=\,0} = -\frac{\al}{r_+}$.}
\begin{equation}\label{eq:static limit of D}
\left. \Ds\right|_{a\,=\,0} = \ETH'\ETH+\ETH\ETH' + 2\,\re(\Psi_2)\;.
\end{equation}
As could be anticipated from the Type~D character of the background, we see that it is part of the Weyl tensor, $\re(\Psi_2)$, not the intrinsic curvature, $\re(\mathcal K)$, that appears most naturally in the GHP-covariant form of the shifted $\Phi_{22}$ for generic values of the angular momentum.
\\\\
On the other hand, the intrinsic curvature presents itself when we trade the GHP-covariant derivatives for the 2d Laplacian plus its associated ejecta. We first expand $\ETH f = [\del+2(-1)\beta-2(-1)\beta'^*]f$ and specialize Eq.~(\ref{eq:anticommutator in terms of 2d laplacian}) to $h = -1$. Then we shuffle the terms around using numerical relations like\footnote{It is possible that these relations embody some hidden meaning. But the two sides of Eq.~(\ref{eq:kerr-newman fact 1}) do not transform in the same way under Eq.~(\ref{eq:GHP group}), so we hesitate to dig deeper.}
\begin{equation}\label{eq:kerr-newman fact 1}
\beta'-\beta^*=\tau'\qquad(\text{Kerr-Newman})
\end{equation}
and
\begin{equation}\label{eq:kerr-newman fact 2}
|\beta'|^2-|\beta|^2 = \frac{a^2}{2|R|^4}\qquad(\text{Kerr-Newman})\;.
\end{equation}
In this way we obtain the following alternative form for Eq.~(\ref{eq:result}):
\begin{align}\label{eq:result with 2d laplacian}
\Ds &= \na_{\text{2d}}^{\;2} +  \tfrac{1-\al R}{r}(\tau\!-\!\tau'^*)R^*\,\del' + \tfrac{1-\al R^*}{r}(\tau^*\!-\!\tau')R\,\del \nonumber\\
&+2r\al\left\{ -2\left( \frac{|R|^2}{|R_0|^2}\,\re(\mathcal K) + \frac{2a^2}{|R|^4}\right)+|\tau\!+\!\tau'^*|^2+\left[ 1+r\al\left( \frac{|R|^2}{2r^2}\right)^2\right] |\tau\!-\!\tau'^*|^2 \right\}\;.
\end{align}
We will refer to the coefficient of $f(\ta,\chi)$ in $\hat{\tilde\Phi}_{22}$, encapsulated by the term in Eq.~(\ref{eq:result with 2d laplacian}) without any derivatives, as the ``mass term.'' It is organized in terms of the intrinsic curvature at the horizon [recall Eq.~(\ref{eq:intrinsic curvature at horizon})] 
%
%
and quantities proportional to some power of the angular momentum. Expressed in this way, the mass term reeks of Kaluza-Klein, but we will leave that for another day. 
Regardless, this form shows clearly which terms go to zero as we turn off the rotation.
\\\\
When $a = 0$ (but $Q \neq 0$), we recover the known spherically symmetric answer:\footnote{At $r = r_+$, we have $\left.\re(\mathcal K)\right|_{a\,=\,0} = \frac{1}{2r_+^2}$. Also, when $a = 0$ the delayed angle $\chi$ becomes the ordinary azimuthal angle $\ph$.}
\begin{equation}
\left. \hat{\tilde\Phi}_{22} \right|_{a\,=\,0} = -\,\frac{c}{4r_+^2}\,\del(U)\left( \na_{\text{2d}}^{\;\;2}-\frac{2\al }{r_+}\right)f(\ta,\ph)\;.
\end{equation}
While the geometrical significance of the mass term in Eq.~(\ref{eq:result with 2d laplacian}) eludes us, the physical significance of the overall factor of $\al$ in shockwave geometries has been emphasized by others.\footnote{We thank Douglas Stanford for explaining this to us.} In the extremal limit, which in this case is $a^2+Q^2 = M^2$ and hence $r_- = r_+$, the surface gravity $\al$ goes to zero (as usual), and the entire mass term vanishes. 
\\\\
As far as we know, the first to point this out in the spherically symmetric situation was Sfetsos, who interpreted it as a breakdown of the solution \cite{sfetsos}. The effect was recently revisited by Leichenauer in the context of entanglement between the conformal field theories dual to the asymptotically-AdS generalization of the Reissner-Nordstr\"om black hole \cite{leichenauer}. And in the context of scattering, the vanishing of the mass term in the operator $\Ds$ is what Maldacena and Stanford call the ``$\beta J$ enhancement'' of the amplitude \cite{maldacena_stanford}. 
\\\\
But let us not get ahead of ourselves. In this paper we are concerned exclusively with the single-shockwave geometry and its interpretation within general relativity. The sun will rise tomorrow, and we will have another opportunity to traverse that wormhole.

\section{Discussion}\label{sec:end}

Inspired by 't~Hooft's S-matrix approach to quantum gravity and Kitaev's recent revival thereof, we have generalized the Dray-'t Hooft gravitational shockwave to the Kerr-Newman black hole using the method of spin coefficients.
\\\\
We have not solved the resulting Green's function equation, $\Ds f \propto \del^2(\vec x_\perp)$. Since $\Ds$ is analytic near $a = 0$, we could perturb around the Dray-'t~Hooft integral formula \cite{shockwave}. Or maybe we should expand in spheroidal harmonics, but we would probably have to resort to numerics for anything beyond a rudimentary understanding.\footnote{Dray and 't~Hooft themselves ``have not attempted to perform the integration explicitly'' for their result \cite{shockwave}. Sfetsos, for his part, did elaborate somewhat on his solutions in Appendix~D of his paper \cite{sfetsos}.} On a different tack, we could perturb other backgrounds by shifting the frame: Shockwaves on Kerr-AdS might eventually lead to precise statements about chaos in a putative dual field theory.\footnote{We thank Nick Hunter-Jones for encouragement in this direction.} 
\\\\
We will conclude with a pedantic remark about the effective action for the horizon field. Given a classical equation of motion, we should ask what variational principle could lead to it. Since the Ricci tensor is linear in $f(\ta,\chi)$, our equation of motion is linear in the field, so we might expect a quadratic action. 
\\\\
But the Lagrangian is proportional to the Einstein-Hilbert curvature $\Pi$, which we have already seen is \textit{linear} in $f$. 
What to make of this? Recall that if the ``equation of motion'' is actually a constraint---which in this case it is---then it should be implemented in the calculus of variations by introducing a Lagrange multiplier.
\\\\
Consider a path integral over all classical fields $f(\ta,\chi)$ that satisfy $\Ds f = 0$:\footnote{For the sake of brevity we are only considering the gravitational part of the action. More generally there should be an $f$-independent function on the right-hand side of the constraint.}
\begin{equation}\label{eq:path integral for horizon field}
\Z \equiv \int \mathcal D f\;\del\left(\Ds f\right) = \int\mathcal Df\,\mathcal Df'\;e^{\,i\int d^2x\,f'\Ds f}\;.
\end{equation}
We have used the Fourier representation of the delta function and thereby concocted a classical field $f'$, which serves as a Lagrange multiplier for the equation $\Ds f = 0$. 
\\\\
The argument of the exponential in Eq.~(\ref{eq:path integral for horizon field}) is 't~Hooft's effective action \cite{t_hooft_interpretation}. This straightforward interpretation of the constraint for the horizon field provides a path-integral sense in which the two shockwaves are canonically conjugate variables. 
\\
\begin{center}\textit{Acknowledgments}\end{center}
We thank Jan Willem Dalhuisen, Aaron Zimmerman, David Nichols, Christopher White, Dave Aasen, Nick Hunter-Jones, Alex Rasmussen, Yonah Lemonik, Douglas Stanford, and Saul Teukolsky for insightful discussions at various points in this endeavor. Y.~B. especially thanks Justin Wilson, Leo Stein, and Alexei Kitaev. J.~S. especially thanks Dirk Bouwmeester. Y.~B. is funded by the Institute for Quantum Information and Matter (NSF Grant PHY-1125565) with support from the Simons Foundation (award number 376205). J.~S. is funded by a UCSB Central Fellowship.

\appendix

\section{Signature change}\label{sec:signs}

In this appendix we sail from West to East, scrupulously marking all signs in our wake. Relics will be tagged by overbars.

\subsection{Basic assumptions}

We begin by flipping the signs of both the base space and internal space metrics:
\begin{equation}\label{eq:signature change 1}
g_{\mu\nu} \equiv \z\,\bar g_{\mu\nu}\;,\;\; \eta_{ab} \equiv \z\,\bar\eta_{ab}\;,\;\; \z \equiv -1\;.
\end{equation}
In terms of the corresponding frames, we have $g_{\mu\nu} = \eta_{ab}\, e_\mu^a e_\nu^b$, $\bar g_{\mu\nu} = \bar\eta_{ab}\,\bar e_\mu^a \bar e_\nu^b$, $\eta_{ab} = g_{\mu\nu} e_a^\mu e_b^\nu$, and $\bar\eta_{ab} = \bar g_{\mu\nu} \bar e_a^\mu \bar e_b^\nu$. Defining $e_{a\mu} \equiv \eta_{ab}\, e_\mu^b$ and $\bar e_{a\mu} \equiv \bar\eta_{ab}\,\bar e_\mu^b$, we obtain from Eq.~(\ref{eq:signature change 1}):
\begin{equation}
e^a_\mu e_{a\nu} = \z\,\bar e^a_\mu \bar e_{a\nu}\;,\;\; e^\mu_a e_{b\mu} = \z\,\bar e^\mu_a \bar e_{b\mu}\;.
\end{equation}
Relative to Chandrasekhar \cite{chandrasekhar}, our null vectors $(l^\mu, l'^\mu, m^\mu, m'^\mu)$ will not flip sign, in which case our null forms $(l_\mu, l'_\mu, m_\mu, m'_\mu) \equiv (g_{\mu\nu}\, l^\nu, g_{\mu\nu}\, l'^\nu, g_{\mu\nu}\, m^\nu, g_{\mu\nu}\, m'^\nu)$ will. This is a choice. 
\\\\
From this---with attention to the fact that the basis is \textit{null}---we infer:
\begin{equation}\label{eq:signature change 2}
e_a^\mu \equiv \bar e_a^\mu\;,\;\; e_\mu^a \equiv \bar e_\mu^a\;.
\end{equation}
Neither $e_a^\mu$ nor $e_\mu^a$ flips sign. What does flip sign is the quantity with both indices lowered:
\begin{equation}\label{eq:signature change 3}
e_{a\mu} = \z\,\bar e_{a\mu}\;.
\end{equation}

\subsection{Spin coefficients flip sign}

If we insert the above definitions into $de^a+\w^a_{\;\;b}\wedge e^b = 0$, we will find that the spin connection \textit{with one index up and one index down} does not flip sign:
\begin{equation}
\w^a_{\;\;b} = \bar\w^a_{\;\;b}\;.
\end{equation}
So $\w_{ab}$ \textit{does} flip sign. Unpacking the 1-form index and recognizing that $dx^\mu = d\bar x^\mu$, we find $(\w_\mu)_{ab} = \z\,(\bar\w_\mu)_{ab}$. Recalling Eq.~(\ref{eq:NP definition of gamma}), we conclude that \textit{the spin coefficients flip sign}:
\begin{equation}
\g_{abc} = \z\,\bar\g_{abc}\;.
\end{equation}
Meanwhile, because of Eq.~(\ref{eq:signature change 3}), \textit{the null Cartan equations are the same in either signature}. So Eq.~(\ref{eq:null cartan}) looks exactly the same as Eq.~(4.13.44) in \textit{Spinors and Spacetime} \cite{penrose}.

\subsection{Curvature scalars do not flip sign}

Next up, curvature. Since $\w^a_{\;\;b}$ does not flip sign, neither does $\W^a_{\;\;b} \equiv d\w^a_{\;\;b}+\w^a_{\;\;c}\wedge\w^c_{\;\;b}$:
\begin{equation}
\W^a_{\;\;b} = \bar\W^a_{\;\;b}\;.
\end{equation}
So $\W_{ab}$ \textit{does} flip sign. As an unavoidable consequence, the Riemann tensor in the internal space with all indices down, $R_{abcd} \equiv (\W_{\mu\nu})_{ab}\,e_c^{\,\mu} e_d^{\,\nu}$, flips sign:
\begin{equation}\label{eq:R_abcd flips sign}
R_{abcd} = \z\,\bar R_{abcd}\;.
\end{equation}
It is misleading to simply assert that the Newman-Penrose equations remain fixed upon changing the metric signature, as if it were to follow as night the day. 
\\\\
Crucially, the Weyl scalars are defined from $C_{abcd}$, which in turn is defined from $R_{abcd}$ [recall Eq.~(\ref{eq:C_abcd})]---\textit{this quantity flips sign under a change of signature}:
%
%
%
\begin{equation}\label{eq:weyl flips sign}
C_{abcd} = \z\, \bar C_{abcd}\;.
\end{equation}
Should we fashion an extra sign in the definition of the Weyl scalars to obviate this? No. Beside the sign from Eq.~(\ref{eq:R_abcd flips sign}), there is also an overall sign choice in the \textit{definition} of the curvature scalars---by sheer happenstance, our conventions in Eq.~(\ref{eq:Weyl scalars}) automatically cancel this additional sign compared to the GHP equations as traditionally written \cite{GHP}. 
\\\\
Meanwhile, since $R^a_{\;\; bcd} = \eta^{ae} R_{ebcd}$ and $R_{abcd} = \z\,\bar R_{abcd}$, the Ricci tensor in the internal space does \textit{not} flip sign:
\begin{equation}\label{eq:ricci doesn't flip}
R_{ab} \equiv R^c_{\;\;acb} = \bar R^c_{\;\;acb} \equiv \bar R_{ab}\;.
\end{equation}
For the Ricci scalars in Eq.~(\ref{eq:Ricci scalars}) we do commission a sign relative to the standard references. 
\\\\
The Einstein-Hilbert curvature sprouts yet another sign:
\begin{equation}\label{eq:scalar curvature flips sign}
\eta^{ab} R_{ab} = \z\,\bar\eta^{ab}\bar R_{ab}\;.
\end{equation}
To maintain the sanctity of the GHP equations, we must begrudgingly define
\begin{equation}\label{eq:our pi}
\Pi \equiv -\,\frac{1}{24} \eta^{ab} R_{ab} = -\z\frac{1}{24}\bar\eta^{ab}\bar R_{ab} = +\bar\Pi\;.
\end{equation}
%

\subsection{Extra sign in GHP derivatives}

Before docking we must ensure that the Icelandic runes make sense. 
Consider the GHP derivatives as defined by Penrose and Rindler \cite{penrose}:
\begin{align}\label{eq:GHP covariant derivatives in mostly minus}
&\bar\THORN \equiv \bar D-2h\,\bar\e-2\bar h\,\bar\e^*\;,\;\; \bar\ETH \equiv \bar\del-2h\,\bar\beta+2\bar h\,\bar\beta'^*\;,\nonumber\\
&\bar\THORN' \equiv \bar D'+2h\,\bar\e'+2\bar h\,\bar\e'^*\;,\;\; \bar\ETH' \equiv \bar\del'+2h\,\bar\beta'-2\bar h\,\bar\beta^*\;.
\end{align} 
Explicitly verifying their GHP covariance on a weighted test function, we see that a certain crucial sign emerges as a result of whether $l^\mu l'_\mu = -m^\mu m'_\mu$ is $+1$ or $-1$. It is this sign that determines the extra signs in Eq.~(\ref{eq:GHP covariant derivatives in mostly minus}) relative to those in Eq.~(\ref{eq:GHP covariant derivatives}). 

\bibliographystyle{unsrt}
\bibliography{References}

\begin{thebibliography}{10}

\bibitem{christodoulou_thermo}
D.~Christodoulou.
\newblock Reversible and irreversible transformations in black-hole physics.
\newblock {\em Phys. Rev. Lett.}, 25(22):1596--1597, 1970.

\bibitem{penrose_floyd}
R.~Penrose and R.~M. Floyd.
\newblock Extraction of rotational energy from a black hole.
\newblock {\em Nat. Phys. Sci.}, 229:177--179, 1971.

\bibitem{carter_thermo}
B.~Carter.
\newblock Rigidity of a black hole.
\newblock {\em Nat. Phys. Sci.}, 238:71--72, 1972.

\bibitem{bekenstein_thermo}
J.~D. Bekenstein.
\newblock Black holes and entropy.
\newblock {\em Phys. Rev. D}, 7(8):2333--2346, 1973.

\bibitem{bardeen_carter_hawking}
J.~M. Bardeen, B.~Carter, and S.~W. Hawking.
\newblock The four laws of black hole mechanics.
\newblock {\em Commun. Math. Phys.}, 31:161--170, 1973.

\bibitem{hawking}
S.~Hawking.
\newblock Black hole explosions?
\newblock {\em Nature}, 248:30--31, 1974.

\bibitem{hawking_2}
S.~Hawking.
\newblock Particle creation by black holes.
\newblock {\em Commun. Math. Phys.}, 43:199--220, 1975.

\bibitem{microstates}
A.~Strominger and C.~Vafa.
\newblock Microscopic origin of the {B}ekenstein-{H}awking entropy.
\newblock {\em Phys. Lett. B}, 379:99--104, 1996.

\bibitem{t_hooft_interpretation}
G.~'t~Hooft.
\newblock The black hole interpretation of string theory.
\newblock {\em Nucl. Phys. B}, (1):138--154, 1990.

\bibitem{t_hooft_S_matrix}
G.~'t~Hooft.
\newblock The scattering matrix approach for the quantum black hole, an
  overview.
\newblock {\em Int. J. Mod. Phys. A11}, pages 4623--4688, 1996.

\bibitem{shenker_stanford}
S.~H. Shenker and D.~Stanford.
\newblock Black holes and the butterfly effect.
\newblock {\em JHEP03}, 67, 2014.

\bibitem{stringy_effects}
S.~H. Shenker and D.~Stanford.
\newblock Stringy effects in scrambling.
\newblock {\em JHEP05}, 132, 2015.

\bibitem{kitaev}
A.~Kitaev.
\newblock A simple model of quantum holography.
\newblock {\em Talk given at the Kavli Institute for Theoretical Physics at the
  University of California, Santa Barbara (Feb. 12, 2015; Apr. 7, 2015; May 27,
  2015)}.

\bibitem{maldacena_stanford_yang}
J.~Maldacena, D.~Stanford, and Z.~Yang.
\newblock Conformal symmetry and its breaking in two-dimensional nearly anti-de
  {S}itter space.
\newblock {\em Prog. Theor. Exp. Phys.}, 2016(12):12C104, 2016.

\bibitem{kitaev_suh}
A.~Kitaev and S.~J. Suh.
\newblock The soft mode in the {S}achdev-{Y}e-{K}itaev model and its gravity
  dual.
\newblock {\em JHEP05}, 183, 2018.

\bibitem{maldacena_stanford}
J.~Maldacena and D.~Stanford.
\newblock Remarks on the {S}achdev-{Y}e-{K}itaev model.
\newblock {\em Phys. Rev. D}, 94:106002, 2016.

\bibitem{AdS2_gravity}
A.~Strominger.
\newblock ${A}d{S}_2$ quantum gravity and string theory.
\newblock {\em JHEP01}, 8, 1999.

\bibitem{witten_syk}
E.~Witten.
\newblock An {SYK}-like model without disorder.
\newblock {\em arXiv:1610.09758 [hep-th]}.

\bibitem{shockwave}
T.~Dray and G.~'t~Hooft.
\newblock The gravitational shock wave of a massless particle.
\newblock {\em Nucl. Phys. B}, 253:173--188, 1985.

\bibitem{flat_shockwave}
P.~C. Aichelburg and R.~U. Sexl.
\newblock On the gravitational field of a massless particle.
\newblock {\em General Relativity and Gravitation}, 2(4):303--312, 1971.

\bibitem{cut_and_paste}
R.~Penrose.
\newblock In L.~O'Raifeartaigh, editor, {\em General Relativity: papers in
  honour of J. L. Synge}. Clarendon, Oxford, 1972.

\bibitem{alonso_zamorano}
R.~Alonso and N.~Zamorano.
\newblock Generalized {K}err-{S}child metric for a massless particle on the
  {R}eissner-{N}ordstr\"om horizon.
\newblock {\em Phys. Rev. D}, 35(6):1798--1801, 1987.

\bibitem{sfetsos}
K.~Sfetsos.
\newblock On gravitational shock waves in curved spacetimes.
\newblock {\em Nucl. Phys. B}, 436:721--745, 1995.

\bibitem{kiem_verlinde_verlinde}
Y.~Kiem, H.~Verlinde, and E.~Verlinde.
\newblock Black hole horizons and complementarity.
\newblock {\em Phys. Rev. D}, 52(12):7053--7065, 1995.

\bibitem{polchinski_chaos}
J.~Polchinski.
\newblock Chaos in the black hole ${S}$-matrix.
\newblock {\em arXiv:1505.08108 [hep-th]}.

\bibitem{AMPS}
A.~Almheiri, D.~Marolf, J.~Polchinski, and J.~Sully.
\newblock Black holes: complementarity or firewalls?
\newblock {\em JHEP}, 2:062, 2013.

\bibitem{polchinski_marolf}
D.~Marolf and J.~Polchinski.
\newblock Gauge-gravity duality and the black hole interior.
\newblock {\em Phys. Rev. Lett.}, 111:171301, 2013.

\bibitem{ligo}
B.~P.~Abbott al.
\newblock Observation of gravitational waves from a binary black hole merger.
\newblock {\em Phys. Rev. Lett.}, 116(061102), 2016.

\bibitem{balasin}
H.~Balasin.
\newblock Generalized {K}err-{S}child metrics and the gravitational field of a
  massless particle on the horizon.
\newblock {\em Class Quantum Grav.}, 17:1913--1920, 2000.

\bibitem{taub}
A.~H. Taub.
\newblock Generalized {K}err-{S}child space-times.
\newblock {\em Annals of Physics}, 134:326--372, 1981.

\bibitem{chandrasekhar}
S.~Chandrasekhar.
\newblock {\em The Mathematical Theory of Black Holes}.
\newblock Oxford University Press, April 28 1983.

\bibitem{penrose}
R.~Penrose and W.~Rindler.
\newblock {\em Spinors and space-time}.
\newblock Cambridge University Press, 1987.

\bibitem{NP}
E.~Newman and R.~Penrose.
\newblock An approach to gravitational radiation by a method of spin
  coefficients.
\newblock {\em J. Math. Phys.}, 3(566):566--578, 1962.

\bibitem{GHP}
R.~Geroch, A.~Held, and R.~Penrose.
\newblock A space-time calculus based on pairs of null directions.
\newblock {\em J. Math. Phys.}, 14(874):874--881, 1973.

\bibitem{boyer_lindquist}
R.~H. Boyer and R.~W. Lindquist.
\newblock Maximal analytic extension of the {K}err metric.
\newblock {\em J. Math. Phys.}, 8:265--281, 1967.

\bibitem{newman_janis}
E.~T. Newman and A.~I. Janis.
\newblock Note on the {K}err spinning-particle metric.
\newblock {\em J. Math. Phys.}, 6:915--917, 1965.

\bibitem{sachs_1961}
R.~K. Sachs.
\newblock Gravitational waves in general relativity. {VI}. {T}he outgoing
  radiation condition.
\newblock In {\em Proc. Roy. Soc. 264, No. 1318}, 1961.

\bibitem{szekeres_refraction}
P.~Szekeres.
\newblock On the propagation of gravitational fields in matter.
\newblock {\em J. Math. Phys}, 7:751--761, 1966.

\bibitem{sachs_lecture}
R.~K. Sachs.
\newblock Gravitational radiation.
\newblock In {\em Relativity Groups and Topology. Lectures Delivered at Les
  Houches During the 1963 Session of the Summer School of Theoretical Physics},
  pages 521--562. DeWitt and DeWitt, published by Gordon \& Breach, Science
  Publishers, Inc, 1964.

\bibitem{dirac_lightcone}
P.~A.~M. Dirac.
\newblock Forms of relativistic dynamics.
\newblock {\em Rev. Mod. Phys.}, 21(3):392--399, 1949.

\bibitem{smarr_E3}
L.~Smarr.
\newblock Surface geometry of charged rotating black holes.
\newblock {\em Phys. Rev. D}, 7(2):289--295, 1973.

\bibitem{szekeres}
P.~Szekeres.
\newblock The gravitational compass.
\newblock {\em J. Math. Phys.}, 6(9):1387--1391, 1965.

\bibitem{petrov}
A.~Z. Petrov.
\newblock The classification of spaces defining gravitational fields.
\newblock {\em General Relativity and Gravitation}, 32(8):1665--1685, 2000.
\newblock [{T}his is an updated version of: {A}. {Z}. {P}etrov. {O}n spaces
  defining gravitational fields. \textit{{D}okl. {A}kad. {N}auk} {SSSR}, {XXXI}
  (1951) 149-152].

\bibitem{griffiths}
J.~B. Griffiths.
\newblock {\em Colliding Plane Waves in General Relativity}.
\newblock Clarendon Press, Oxford, (1991); Dover reprint 2016.

\bibitem{rides_again}
M.~Fels and A.~Held.
\newblock Kerr-{S}child rides again.
\newblock {\em General Relativity and Gravitation}, pages 61--68, 1989.

\bibitem{matzner}
R.~A. Matzner.
\newblock Behavior of ray optics in the {D}ray-'t {H}ooft geometry.
\newblock {\em Nucl. Phys. B}, 266:661--668, 1986.

\bibitem{o'neill}
B.~O'Neill.
\newblock {\em The Geometry of Kerr Black Holes}.
\newblock A K Peters, Ltd., Wellesley, Massachusetts, 1995.

\bibitem{leichenauer}
S.~Leichenauer.
\newblock Disrupting entanglement of black holes.
\newblock {\em Phys. Rev. D}, 90(046009), 2014.

\end{thebibliography}
\end{document}